\documentclass[entropy,article,submit,moreauthors,pdftex]{Definitions/mdpi} 

\firstpage{1} 
\pubvolume{1}
\issuenum{1}
\articlenumber{0}
\pubyear{2021}
\copyrightyear{2020}
\datereceived{} 
\dateaccepted{} 
\datepublished{} 
\hreflink{https://doi.org/} % If needed use \linebreak
\pdfoutput=1

\usepackage{amssymb}
\usepackage[active]{srcltx} % this should allow forward search
\usepackage{mathrsfs}
\usepackage{Milancommands}
  \usetikzlibrary {calc,positioning,graphs,quotes,arrows.meta,math}

\newcommand{\rin}{R_{\text{ITL}}^{n}}

\newcommand{\Rev}{\mathscr{R}}
\newcommand{\Qev}{\mathscr{Q}}
\newcommand{\Uev}{\mathscr{U}}
\newcommand{\Yev}{\mathscr{Y}}
\newtheorem{thm}{\textbf{Theorem}}
\newtheorem{defn}{\textbf{Definition}}

\newtheorem{lem}{Lemma}
\newtheorem{rem}{Remark}

\newtheorem{exple}{Example}
\tikzset{
>={Latex[width=3pt,length=4pt]},
 bloq/.style={rectangle, rounded corners=.7mm, inner ysep=1.3pt, minimum width=5mm, draw=black, font=\itshape},
 sm/.style={circle, draw=black,inner sep=.7pt},
 cov/.style 2 args={to path={-- ++(#1,0) node[pos=.5,above] {#2}|- (\tikztotarget)}},
 hv/.style 2 args={to path={-| (\tikztotarget) node[left,pos=#1]{#2}}},
 vh/.style={to path={|- (\tikztotarget)}}
 }
\Title{Directed Data-Processing Inequalities for Systems with Feedback}

\TitleCitation{Directed Data-Processing Inequalities for Systems with Feedback}

\Author{Milan S. Derpich$^{1}$ and Jan {\O}stergaard $^{2}$}

\address{%
$^{1}$ \quad  Department of Electronic Engineering, Universidad T\'ecnica Federico Santa Mar\'ia, Casilla 110-V, Valpara\'iso, Chile; milan.derpich@usm.cl.\\
$^{2}$ \quad Department of Electronic Systems, Aalborg University, Fredrik Bajers Vej 7, DK-9220, Aalborg, Denmark; janoe@ieee.org.}

\abstract{
We present novel data-processing inequalities relating the mutual information and the directed information in systems with feedback.  The internal blocks within such systems are restricted only to be causal mappings, but are allowed to be non-linear, stochastic and time varying.  These blocks can for example represent source encoders, decoders or even communication channels. Moreover, the involved signals can be arbitrarily distributed.
Our first main result relates mutual and directed informations and can be interpreted as a law of conservation of information flow.
Our second main result is a pair of data-processing inequalities (one the conditional version of the other) between nested pairs of random sequences entirely within the closed loop.
Our third main result is introducing and characterizing the notion of in-the-loop (ITL) transmission rate for channel coding scenarios in which the messages are internal to the loop. Interestingly, in this case the conventional notions of transmission rate associated with the entropy of the messages and  of channel capacity based on maximizing the mutual information between the messages and the output turn out to be inadequate. 
 Instead, as we  show, the ITL transmission rate is the unique notion of rate for which a channel code attains zero  error probability if and only if  such ITL rate does not exceed the corresponding directed information rate from messages to decoded messages.
We apply our data-processing inequalities to show that the supremum of   achievable (in the usual channel coding sense) ITL transmission rates is upper bounded by the supremum of the directed information rate across the communication  channel. Moreover, we present an example in which this upper bound is attained.
Finally, we further illustrate the applicability of our results by discussing how they make possible the generalization of two fundamental inequalities known in networked control literature.
}


\begin{document}
\end{paracol}
%%%%%%%%%%%%%%%%%%%%%%%%%%%%%%%%%%%%%%%%%%
\section{Introduction}\label{sec:intro}
The data-processing inequality states that, if $\rvax,\rvay,\rvaz$ are random variables such that $\rvax$ and $\rvaz$ become independent when conditioning upon $\rvay$, then \begin{align}
I(\rvax;\rvay                                                                                                                                                                           )
\ge I(\rvax;\rvaz)
\\
I(\rvay;\rvaz)\geq I(\rvax;\rvaz),
\end{align}
where $I(\rvax;\rvay)$ denotes the mutual information between $\rvax$  and $\rvay$~\cite[p.~252]{covtho06} (a definition of mutual information is provided in Section~\ref{subsec:mi} below).
Among its many uses, the data-processing-inequality plays a key role in the proof of the converse part (i.~e., outer bounds) in
rate-distortion~\cite{salcad19,lintsu18,yangro17,derost12}\cite[p.~317]{covtho06},
channel capacity~\cite{ramite21,sonzha20,makur-20}\cite[pp.~208, 217, 540 and~566]{covtho06},
and joint source-channel coding theorems~\cite{kosver13,huanar12,stemer06}~\cite[p.~221]{covtho06}.
% \cite{sahmit06}

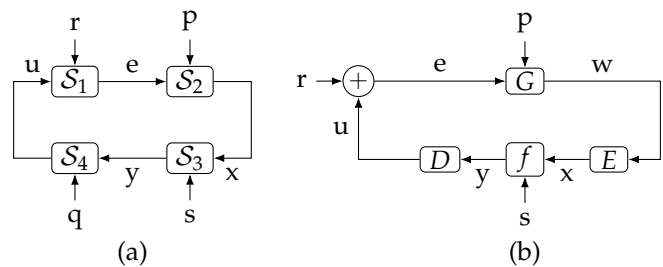
\begin{figure}[htbp]
\centering
\begin{tikzpicture}[node distance=9mm]
 \def\la{3.5mm}
\node (b1) [bloq] {$\Ssp_{1}$};
\node (b2) [bloq, right =of b1] {$\Ssp_{2}$};
\node (b3) [bloq, below = of b2, yshift=3mm] {$\Ssp_{3}$};
\node (b4) [bloq, left =of b3] {$\Ssp_{4}$};

\path (b1) edge[->] node[pos=0.5, above] {$\rvae$} (b2);
\path (b3) edge[->] node[pos=0.5, below] {$\rvay$} (b4);
\draw[->] (b2.east) --++ (5mm,0) |-($(b3.east) $) node[pos=.75,below] {$\rvax$};
\draw[->] (b4.west) --++(-.5,0) |-(b1.west) node[ pos=.75, above] {$\rvau$};

\draw[<-] (b1.north)--++(0,\la) node[above] {$\rvar$};
\draw[<-] (b2.north)--++(0,\la) node[above] {$\rvap$};
\draw[<-] (b3.south)--++(0,-\la) node[below] {$\rvas$} coordinate (ps);
\draw[<-] (b4.south)--++(0,-\la) node[below] {$\rvaq$} coordinate (pq);
\node  at ($.5*(ps)+.5*(pq)-(0,2*\la)$) {(a)};
\end{tikzpicture}
\hspace{.3cm}
\begin{tikzpicture}[node distance=9mm]
\vspace{-1cm}
 \def\la{3.5mm}
\matrix[row sep=.6cm, column sep=.6cm]{
\node (su) [sm] {\small{$+$}}; &
                                                    &
\node (G) [bloq] {$G$};             &\\
                                                     &
\node (D) [bloq] {$D$};             &
\node (f) [bloq] {$f$};                &
\node (E) [bloq] {$E$};\\
};
\graph{
(su) -> ["$\rvae$"] (G) ->[cov={18mm}{$\rvaw$}] (E) ->["$\rvax$"]  (f) ->["$\rvay$"]  (D) ->[hv={.75}{$\rvau$}] (su);
};
\draw[<-] (su.west)--++(-\la,0) node[left] {$\rvar$};
\draw[<-] (G.north)--++(0,\la) node[above] {$\rvap$};
\draw[<-] (f.south)--++(0,-\la) node[below] {$\rvas$} coordinate (pb);
\node at ($(pb)-(0,1.9*\la)$) {(b)};
 \end{tikzpicture}
\caption{(a): The general system considered in this work. 
(b): A special case of (a), corresponding to the closed-loop system studied in~\cite{mardah05}.} 
\label{fig:diagramas}
\end{figure}

It is well known that the mutual information has an important limitation in systems with feedback, such as  the one shown in Fig.~\ref{fig:diagramas}-(a).
In this system, 
$\rvap$,
$\rvaq$,
$\rvar$,
$\rvas$,
$\rvae$,
$\rvau$,
$\rvax$ and
$\rvay$ are random sequences and the blocks 
$\Ssp_{1},\ldots,\Ssp_{4}$
are causal mappings with an added delay of at least one sample.
As  pointed out in~\cite{massey90}, for sequences inside the loop, such as $\rvax$ and $\rvay$, $I(\rvax;\rvay)$ does not distinguish the probabilistic interdependence produced by the effect $\rvax$ has on $\rvay$ from that stemming from the influence of $\rvay$ on $\rvax$.
This limitation motivated the introduction of the directed information in~\cite{massey90}.
This notion  assesses the amount of information that causally ``flows'' from a given random and ordered sequence to another. 
For this reason, it has increasingly found use in diverse applications, from
characterizing the capacity of channels with feedback~\cite{massey90,kramer98,tatmit09,li-eli11},
the rate distortion function under causality constraints~\cite{derost12},
establishing some of the fundamental limitations in networked control~\cite{tatiko00,mardah05,mardah08,silder11,silder10,silder11b,tanesf18}, 
determining causal relationships in neural networks~\cite{quicol11},
to 
portfolio theory and hypothesis testing~\cite{perkim11}, to name a few.

The directed information from a random%
\footnote{
Hereafter we use non-italic letters (such as $\rvax$) for random variables, denoting a particular realization by the corresponding italic character, $x$.
}
 sequence $\rvax^{k}$ to a random sequence  $\rvay^{k}$ is defined as 
\begin{align}\label{eq:directed_inf_def}
 I(\rvax^{k}\to\rvay^{k}) \eq \Sumfromto{i=1}{k}I(\rvay(i);\rvax^{i}|\rvay^{i-1}),
\end{align}
where the notation $\rvax^{i}$ represents the sequence $\rvax(1),\rvax(2),\ldots, \rvax(i)$ 
and $I(\rvax;\rvay|\rvaz)$ is the mutual information between $\rvax$ and $\rvay$ conditioned on (or given) $\rvaz$.
The causality inherent in this definition becomes evident when comparing it with the mutual information between $\rvax^{k}$ and $\rvay^{k}$, given by $I(\rvax^{k};\rvay^{k})=\sumfromto{i=1}{k}I(\rvay(i);\rvax^{k}|\rvay^{i-1})$.
In the latter sum, what matters is the amount of information about the \textit{entire} sequence $\rvax^{k}$
present in $\rvay(i)$, given the past values $\rvay^{i-1}$.
By contrast, in the conditional mutual informations in the sum of~\eqref{eq:directed_inf_def}, only the past and current values of $\rvax^{k}$ are considered, that is, $\rvax^{i}$.
Thus,  $I(\rvax^{k}\to\rvay^{k})$ represents the amount of information causally conveyed from $\rvax^{k}$ to $\rvay^{k}$.
A related notion is the causally conditioned directed information introduced in~\cite{kramer98},  defined as
\begin{align}
I(\rvax^{k}\to \rvay^{k}\parallel\rvaq^{k})
\eq 
\Sumfromto{i=1}{k}I(\rvay(i);\rvax^{i}|\rvay^{i-1},\rvaq^{i})
\end{align}

In this paper, we derive inequalities involving directed and mutual informations within feedback systems.
For this purpose, we consider the general feedback system shown in Fig.~\ref{fig:diagramas}-(a).
In this diagram, the blocks $\Ssp_{1},\ldots, \Ssp_{4}$ represent possibly non-linear and time-varying causal discrete-time systems such that the total delay of the loop is at least one sample. 
These blocks can model, for example, source encoders, decoders or even communication channels.
In the same figure, $\rvar,\rvap,\rvas,\rvaq$ are exogenous random signals (scalars, vectors or sequences), which could represent, for example, any combination of disturbances, noises, random initial states or side informations.
We note that any of these exogenous signals, in combination with its corresponding deterministic mapping $\Ssp_{i}$, can also yield any desired stochastic causal mapping (for example, a noisy communication channel, a zero-delay source coder or decoder, or a causal dynamic system with disturbances and a random initial state).

\subsection{Main Contributions}
Our \textbf{first two main results} are the following theorems.

\begin{figure}[htpb]
 \centering
\input{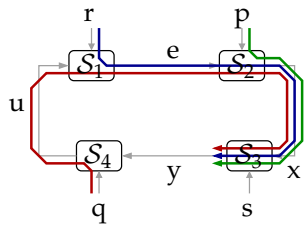}
\caption{The flow of information between exogenous signals $(\rvap,\rvaq,\rvar)$ and the internal signal $\rvay$ equals the directed information from $\rvax^{k}$ to $\rvay^{k}$ when $\rvas\Perp(\rvap,\rvaq,\rvar)$.}
\label{fig:information_flow}
\end{figure}

The first theorem states a fundamental result, \textit{which relates the directed information between two signals within a feedback loop, say $\rvax$ and $\rvay$, to the mutual information between an external set of signals and $\rvay$}:
\begin{thm}\label{thm:main}
\textit{ In the system shown in Fig.~\ref{fig:diagramas}-(a), it holds that
\begin{align}\label{eq:main_thm}
 I(\rvax^{k}\to \rvay^{k}) 
=
 I(\rvaq^{k},\rvar^{k},\rvap^{k}\to\rvay^{k})
- 
I(\rvaq^{k},\rvar^{k},\rvap^{k}\to\rvay^{k}\parallel \rvax^{k})
\leq I(\rvap^{k},\rvaq^{k},\rvar^{k}\,;\rvay^{k}), \fspace \forall k\in\Nl,
\end{align}
with equality achieved if $\rvas$ is independent of $(\rvap,\rvaq,\rvar)$. }
\finenunciado
\end{thm}
The proof is in Section~\ref{sec:prfmain}.
This fundamental result, which for the cases in which%
\footnote{Here, and in the sequel, we use the notation $\rvax \Perp\rvay $ 
to mean ``$\rvax$ is independent of $\rvay$''.}
$\rvas\Perp(\rvap,\rvaq,\rvar)$ can be understood as a \textit{law of conservation of information flow}, is illustrated in Fig.~\ref{fig:information_flow}.
For such cases, the information causally conveyed from $\rvax$ to $\rvay$ equals the information flow from $(\rvaq,\rvar,\rvap)$ to $\rvay$.   
When $(\rvap,\rvaq,\rvar)$ are not independent of $\rvas$, part of the mutual information between $(\rvap,\rvaq,\rvar)$ and $\rvay$ 
(corresponding to the term $I(\rvaq^{k},\rvar^{k},\rvap^{k}\to\rvay^{k}\parallel \rvax^{k})$) 
can be thought of as being ``leaked'' through $\rvas$, thus bypassing the forward link from $\rvax$ to $\rvay$.
This provides an intuitive interpretation for~\eqref{eq:main_thm}. 

\begin{rem}
Theorem~\ref{thm:main} implies that $I(\rvax^{k}\to \rvay^{k})$ is only a part of (or at most equal to) the  information ``flow'' between all the exogenous signals entering the loop outside the link $\rvax\to \rvay$ 
(namely $(\rvaq,\rvar,\rvap)$), and $\rvay$.
In particular, if $(\rvap,\rvaq,\rvar)$ were deterministic, then $I(\rvax^{k}\to \rvay^{k})=0$, regardless of the blocks $\Ssp_{1},\ldots,\Ssp_{4}$ and irrespective of the nature of $\rvas$. 
\finenunciado
\end{rem}

Our second main result is the following theorem, which relates directed informations involving four different sequences internal to the loop.
The proof is in Appendix~\ref{sec:proofs} on page~\pageref{prf:full_D-DPI}.
\begin{thm}[Full Closed-Loop Directed Data-Processing Inequality]\label{thm:full_D-DPI}
Consider  the system shown in Fig.~\ref{fig:diagramas}-(a).
\begin{enumerate}
 \item 
 If  $(\rvaq,\rvas)\Perp (\rvar,\rvap)$ and $\rvaq\Perp\rvas$,
 or
 if  $(\rvap,\rvas)\Perp (\rvar,\rvaq)$ and $\rvap\Perp\rvas$,
then
\begin{align}
 I(\rvax^{k}\to \rvay^{k})
 \geq 
I(\rvae^{k}\to \rvau^{k}).\label{eq:full_Dpi}
\end{align}
\item If $(\rvaq,\rvas)\Perp (\rvar,\rvap)$ and%
 \footnote{The Markov chain notation $\rvaa\leftrightarrow \rvab \leftrightarrow \rvac$ means ``$\rvaa$ and $\rvac$ are independent when $\rvab$ is given''.}
 $\rvaq_{i+1}^{k}\leftrightarrow \rvaq^{i} \leftrightarrow \rvas^{i}$ for $i=1,2,\ldots,k-1$, then
\begin{align}\label{eq:silderok}
 I(\rvax^{k}\to\rvay^{k}\Vert\rvaq^{k})\geq I(\rvae^{k}\to\rvau^{k}).
\end{align}
\end{enumerate}
\finenunciado
\end{thm}
To the best of our knowledge, Theorem~\ref{thm:full_D-DPI} is the first result available in the literature 
providing a lower bound to the gap between two nested directed informations, involving four different signals inside the feedback loop.
This result can be seen as the first full extension of the open-loop (traditional) data-processing inequality, to  arbitrary closed-loop scenarios. 
(Notice that there is no need to consider systems with more than four mappings, since all external signals entering the loop between a given pair of internal signals can be regarded as exogenous inputs to a single equivalent deterministic mapping.)

Our third main contribution is introducing the notion of 
\emph{In-the-loop} (ITL) transmission rate (in Section~\ref{sec:itl}) for the (seldom considered) channel-coding scenario in which the messages to be transmitted and the communication channel are internal to a feedback loop.
We show that the supremum of the directed information rate across such channel upper bounds the achievable ITL transmission rates.
Moreover, we present an example in which this upper bound is attainable.
This gives further operational meaning to the directed information rate in closed-loop scenarios.

Finally, we provide additional examples of  the applicability of our results by discussing
how they allow one 
to obtain the generalizations of two fundamental inequalities known in networked control literature.
The first one  appears in~\cite[Lemma~4.1]{mardah05} and is written in~\eqref{eq:mardah_dir_minus_mutual} below.
This generalization is a consequence of Theorem~\ref{thm:three_full_loops} and is discussed in remarks~\ref{rem:maruno} and~\ref{rem:mardos} below.
The second generalization applies to~\cite[Theorem~4.1]{silder11} and is described on page~\pageref{pag:silder11} below.
It is an application of Theorem~\ref{thm:full_D-DPI} that has just been carried out  by the authors in~\cite{derost21b}, all the more important  since, as we also revealin that note, there is a flaw in the proof of~\cite[Theorem~4.1]{silder11}.

\subsection{Existing Related Results}
There exist several results characterizing the relationship between $I(\rvax^{k}\to\rvay^{k})$ and $I(\rvax^{k};\rvay^{k})$.
First, it is well known that $I(\rvax^{k}\to\rvay^{k})\leq I(\rvax^{k};\rvay^{k})$, with equality if and only if $\rvay^{k}$ is causally related to $\rvax^{k}$% 
% and there is no feedback from $\rvay^{k}$ to $\rvax^{k}$
~\cite{massey90}.
A conservation law of mutual and directed information has been found in~\cite{masmas05}, which asserts that 
$I(\rvax^{k}\to\rvay^{k}) + I(0*\rvay^{k-1}\to \rvax^{k}) = I(\rvax^{k};\rvay^{k})$, where $0*\rvay^{k-1}$ denotes the concatenation $0,\rvay(1),\ldots,\rvay^{k-1}$.

Given its prominence in settings involving feedback, it is perhaps in these scenarios where the directed information becomes most important.
For instance, the directed information has been instrumental in characterizing the capacity of channels with feedback (see, e.g.,~\cite{tatmit09,kim-yh08,li-eli11} and the references therein), as well as the rate-distortion function in setups involving feedback~\cite{zamkoc08,silder11,silder11b,silder10,derost12}.

For the simple case in which all the systems $\set{\Ssp_{i}}_{i=1}^{4}$ are linear time invariant (LTI) and stable, and assuming $\rvap,\rvax,\rvaq=0$ (deterministically), it was shown in~\cite{zhasun06} that $I(\rvar^{k}\to \rvae^{k})$ does not depend on whether there is feedback from $\rvae$ to $\rvau$ or not.

Inequalities between mutual and directed informations in a less restricted setup, shown in Fig.~\ref{fig:diagramas}-(b), have been found in~\cite{mardah05,mardah08}.  
In that setting (a networked-control system), $G$ is a strictly causal LTI dynamic system having 
(vector) state sequence $\set{\rvex(i)}_{i=0}^{\infty}$, with 
$\rvap\eq\rvex(0)$ being the random initial state in its state-space representation.
The external signal $\rvar$ (which could correspond to a disturbance) is statistically independent of $\rvas$, the latter corresponding to, for example, side information or channel noise.
Both are also statistically independent of $\rvap$.

The blocks labeled $E$, $D$ and $f$ correspond to an encoder, a decoder and a channel, respectively, all of which are causal.
The channel $f$ maps $\rvas^{k}$ and $\rvax^{k}$ to $\rvay(k)$ in a possibly time-varying manner, i.e.,
$
 \rvay(k)=f(k,\rvax^{k},\rvas^{k}).
$
Similarly, the concatenation of the encoder, the channel and the decoder, maps $\rvas^{k}$ and $\rvaw^{k}$ to $\rvau(k)$ as a possibly time-dependent function
$
 \rvau(k)=\psi(k,\rvaw^{k},\rvas^{k}).
$
Under these assumptions, the following fundamental result was shown in~\cite[Lemma~5.1]{mardah08}:
\begin{align}\label{eq:Martins_first}
 I(\rvar^{k},\rvap\,;\,\rvau^{k})
\geq 
I(\rvar^{k}; \rvau^{k})
+
I(\rvap;\rvae^{k}).
\end{align}
% \end{thm}
%
By further assuming in~\cite{mardah08} that the decoder $D$ in Fig.~\ref{fig:diagramas}-(b) is deterministic,
the following Markov chain naturally holds,
\begin{align}\label{eq:MC_martins}
 (\rvap,\rvar^{k}) 
\longleftrightarrow
% e^{k}
% \longleftrightarrow
% x^{k}
% \longleftrightarrow
\rvay^{k}
\longleftrightarrow
\rvau^{k},
\end{align}
leading directly to 
%
% \begin{coro}[{From~\cite[Corollary~5.3]{mardah08}}]\label{coro:martins5.3}
%  If $r$, $s$ and $\bx_{0}$ are mutually independent and if decoder $D$ in Fig.~\ref{fig:diagramas} is deterministic, it holds that
\begin{align}\label{eq:la_de_martins08}
  I(\rvar^{k},\rvap\,;\,\rvay^{k})
\geq 
I(\rvar^{k}; \rvau^{k})
+
I(\rvap;\rvae^{k}),
\end{align}
% \end{coro}
which is found in the proof of~\cite[Corollary~5.3]{mardah08}.
The deterministic nature of the decoder $D$ played a crucial role in the proof of this result,  since otherwise the Markov chain~\eqref{eq:MC_martins} does not hold, in general, due to the feedback from $\rvau$ to $\rvay$. 

Notice that both~\eqref{eq:Martins_first} and~\eqref{eq:la_de_martins08} provide lower bounds to mutual information as the sum of two mutual information terms, each of them relating a signal \textit{external} to the loop (such as $\rvap,\rvar^{k}$) to a signal \textit{internal} to the loop (such as $\rvau^{k}$ or $\rvay^{k}$).
Instead, the inequality 
\begin{align}\label{eq:la_de_Massey}
I(\rvax^{k}\to\rvay^{k})\geq I(\rvar^{k};\rvay^{k}),
\end{align}
which holds for the system in Fig.~\ref{fig:diagramas}-(a) and appears 
in~\cite[Theorem~3]{massey90} (and rediscovered later in~\cite[Lemma~4.8.1]{tatiko00}),
involves the directed information between two internal signals and the mutual information between the second of these and an external sequence.

\begin{rem}
By using~\eqref{eq:chainrule_I}, 
$
I(\rvap^{k},\rvaq^{k},\rvar^{k};\rvay^{k})
=
I(\rvar^{k};\rvay^{k})
+
I(\rvap^{k},\rvaq^{k};\rvay^{k}|\rvar^{k})
$.
Then, applying 
Theorem~\ref{thm:main}, we recover~\eqref{eq:la_de_Massey}, whenever $\rvas\Perp(\rvaq,\rvar,\rvap)$.
Thus,~\cite[Theorem~3]{massey90} and~\cite[Lemma~4.8.1]{tatiko00}) can be obtained as a corollary of Theorem~\ref{thm:main}.
\finenunciado
\end{rem}

A related bound,  similar to~\eqref{eq:la_de_martins08} 
but involving information rates and with the leftmost mutual information 
replaced by the directed information from $\rvax^{k}$ to $\rvay^{k}$ (which are two signals internal to the loop), has been obtained in~\cite[Lemma~4.1]{mardah05} for the networked control system of Fig.~\ref{fig:diagramas}-(b):
\begin{align}\label{eq:mardah_dir_minus_mutual}
 \bar{I}(\rvax\to \rvay)  \geq \bar{I}(\rvar;\rvau)+ \lim_{k\to\infty}\frac{I(\rvap;\rvae^{k})}{k},
\end{align}
with $\bar{I}(\rvax\to \rvay)\eq \lim_{k\to\infty}\frac{1}{k}I(\rvax^{k}\to \rvay^{k})$
and
$\bar{I}(\rvar;\rvau)\eq \lim_{k\to\infty}\frac{1}{k}I(\rvar^{k};\rvau^{k})$,
provided $\sup_{i\geq 0}\Expe{\rvex(i)^{T}\rvex(i)}<\infty$.
This result relies on
three assumptions: 
a) that the channel $f$ is memory-less and satisfies a ``conditional invertibility'' property, b) a finite-memory condition, and c) a fading-memory condition, these two related to the decoder $D$ (see Fig.~\ref{fig:diagramas}).  

It is worth noting that, as defined in~\cite{mardah05}, these assumptions upon $D$ exclude the use of side information by the decoder and/or the possibility of $D$ being affected by random noise or having a random internal state which is non-observable (please see~\cite{mardah05} for a detailed description of these assumptions). 

\begin{rem}\label{rem:maruno}
 In Section~\ref{sec:mad} we present Theorem~\ref{thm:three_full_loops}, which yields~\eqref{eq:mardah_dir_minus_mutual} as a special case, but for the general system of Fig.~\ref{fig:diagramas}-(a) and with no other assumption than mutual independence between $\rvar,\rvap,\rvaq,\rvas$.
 Moreover,  since with this independence condition Theorem~\ref{thm:main} yields $I(\rvax^{k}\to\rvau^{k})=I(\rvar^{k},\rvap^{k};\rvau^{k})$, the same happens with~\eqref{eq:Martins_first}.
 \finenunciado
\end{rem}

The inequality~\eqref{eq:la_de_Massey} has  been extended in~\cite[Theorem~1]{li-eli11}, for the case of discrete-valued random variables and assuming $\rvas\Perp(\rvar,\rvap,\rvaq)$, as the following identity (written in terms of the signals and setup shown in Fig.~\ref{fig:diagramas}-(a)):
\begin{align}\label{eq:latest1}
 I(\rvax^{k}\to\rvay^{k})
 =
 I(\rvap^{k},\rvay^{k})
 +
 I(\rvax^{k}\to\rvay^{k}|\rvap^{k} ).
\end{align}
Letting $\rvaq=\rvas$ in Fig.~\ref{fig:diagramas}-(a) and with the additional assumption that $(\rvap,\rvas)\Perp\rvaq$, it was also shown in~\cite[Theorem~1]{li-eli11} that
\begin{align}\label{eq:latest2}
 I(\rvax^{k}\to\rvay^{k})
 =
 I(\rvap^{k};\rvay^{k})
 +
 I(\rvaq^{k-1};\rvay^{k})
+
 I(\rvap^{k};\rvaq^{k-1}|\rvay^{k}),
\end{align}
for the cases in which $\rvau(i)=\rvay(i)+\rvaq(i)$ (i.e., when the concatenation of $\Ssp_{4}$ and $\Ssp_{1}$ corresponds to a summing node).
In~\cite{li-eli11},~\eqref{eq:latest1} and~\eqref{eq:latest2} play important roles in characterizing the capacity of channels with noisy feedback.

To the best of our knowledge,%
~\eqref{eq:Martins_first},~\eqref{eq:la_de_martins08},~\eqref{eq:la_de_Massey}~\eqref{eq:mardah_dir_minus_mutual},~\eqref{eq:latest1} and~\eqref{eq:latest2}
are the only results available in the literature which lower bound the difference between an internal-to-internal directed information and an external-to-internal mutual information.
There exist even fewer published results in relation to inequalities between two directed informations involving only signals internal to the loop.
To the best of our knowledge, the only inequality of this type in the literature is the one found in the proof of 
Theorem~4.1 of~\cite{silder11}.\label{pag:silder11}
The latter takes the form of a (conditional) data-processing inequality for directed informations in closed-loop systems, and states that
\begin{align}\label{eq:silder11}
I(\rvax^{k}\to \rvay^{k}\parallel\rvaq^{k})\geq  I(\rvax^{k}\to \rvau^{k}),                                
\end{align}
provided
 $\rvaq\Perp(\rvar,\rvap)$
and if $\Ssp_{4}$ is such that $\rvay^{i}$ is a function of $(\rvau^{i},\rvaq^{i})$ (i.e., if $\Ssp_{4}$ is conditionally invertible) $\forall i$.
      
\begin{figure}[htbp]
\centering
\begin{tikzpicture}[node distance=9mm]
 \def\la{3.5mm}
\node (b1) [bloq] {$\Fsp$};
\node (b3) [bloq, below = of b1, yshift=3mm, xshift=10mm] {$\Esp$};
\node (b4) [bloq, below =of b1, yshift=3mm,xshift= -10mm] {$\Dsp$};

\draw[->] (b1.east) --++ (18mm,0) coordinate (c) |-($(b3.east) $) node[pos=.75,below] {$\rvax$};
\draw[->] (b3.west) -- (b4.east)node[above,pos=.5]  {$\rvay$};
\draw[->] (b4.west) --++(-8mm,0) |-(b1.west) node[ pos=.75, above] {$\rvau$};

\draw[<-] ($(b1.north)-(.12,0)$) --++(0,\la) node[above ] {$\rvar$} coordinate (a);
\draw[<-] ($(b1.north)+(.12,0)$) --++(0,\la) node[above =-.08] {$\rvap$};
\draw[<-] (b3.south)--++(0,-\la) node[below] {$\rvas$} coordinate (ps);
\draw[<-] (b4.south)--++(0,-\la) node[below] {$\rvaq$} coordinate (pq);

\def\gap{1.5mm}
\coordinate (h) at (\gap,0);
\coordinate (v) at (0,\gap);
\coordinate (b) at ($0.5*(b1.east)+0.5*(b1.west)$);
\coordinate (cc) at (b3.east);
\coordinate (f) at ($(b4.west)-(2.1,0)$);

\end{tikzpicture}
\caption{The networked control system considered in \cite[Fig.~2]{silder11}, slightly   simplified.
The variables $\rvar,\rvap,\rvax,\rvay$ correspond to $d,x_{o},y,s$ in ~\cite{silder11}, respectively.
}\label{fig:silder}
\end{figure}
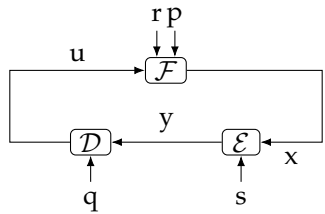

Inequality~\eqref{eq:silder11} plays a crucial role in~\cite{silder11}, since it allowed~\cite[Thm.~4.1]{silder11} to lower bound the average data rate across a digital error-free channel by a directed information. 
% (In~\cite{silder11}, $\rvaq$ corresponded to a random dither signal in an entropy-coded dithered quantizer.) 
The setup considered in that theorem is shown in Fig.~\ref{fig:silder}, where $\Fsp$ is a plant, and $\Esp$, $\Dsp$ are (source) encoder and decoder, respectively.
In this figure, the variables have been adapted to match those in Fig.~\ref{fig:2systems}-(a) ($\rvar,\rvap, \rvax$ correspond to disturbance, initial state and plant output, respectively). 
Assuming $(\rvar,\rvap)\Perp(\rvaq,\rvas)$ and a conditionally invertible decoder,  and letting $R(i)$ be the expected length (in bits) necessary for a binary representation of $\rvay(i)$ given $\rvaq^{i}$, it states that
$ \frac{1}{k} \sumfromto{i=1}{k} R(i)\geq \frac{1}{k} I(\rvax^{k}\to\rvau^{k}), \; k=1,2,\ldots.$
This is a key result, because, combined with~\cite[eq.~(9)]{silder11}, 
it yields
\begin{align}\label{eq:sdw}
  \frac{1}{k} I(\rvax^{k}\to\rvau^{k})
  \leq\frac{1}{k} \Sumfromto{i=1}{k} R(i)\leq \frac{1}{k} I(\rvax^{k}\to\rvau^{k})+1 \fspace\text{[bits/sample]}, \fspace k=1,2,\ldots.
\end{align}
This result highlights the operational meaning of the directed information as a lower bound (tight to within one bit) to the data rate of any given source code in a closed-loop system.
This fact has been a crucial ingredient in characterizing the best 
rate-performance achievable in Gaussian linear quadratic networked 
control~\cite{silder16,tanesf18}, demonstrating the relevance of directed data-processing inequalities.

Unfortunately, as we will reveal in~\cite{derost21b}, the proof of~\cite[Theorem 4.1]{silder11}  turns out to be invalid, since it relies upon~\cite[Lemma 4.2]{silder11}, whose first claim does not hold.
In~\cite{derost21b} we use Theorem~\ref{thm:full_D-DPI} to prove Theorem~4.1 of~\cite{silder11} without requiring a conditionally invertible decoder.
This further illustrates the applicability of our results.

In~\cite[Lemma~1]{tanesf18} another  data-processing inequality is stated, which for the system in Fig.~\ref{fig:diagramas}-(a) is equivalent to
\begin{align}\label{eq:tana}
 I(\rvax^{k}\to\rvay^{k}\Vert\rvau_{+}^{k-1})\geq I(\rvax^{k}\to \rvau^{k}), \fspace k=1,2,\ldots,
\end{align}
where
$I(\rvax^{k}\to\rvay^{k}\Vert\rvau_{+}^{k-1})\eq \sumfromto{i=1}{k}I(\rvax^{i};\rvay(i)|\rvay^{i-1},\rvau^{i-1})$.
However, in~\cite{tanesf18} the blocks $\Ssp_{3},\ \Ssp_{4}$ are defined implicitly, writing instead their input-output relation as collections of stochastic kernels
$\mathbb{P}(\rvay(i)|\rvax^{i},\rvay^{i-1})$,
$\mathbb{P}(\rvau(i)|\rvay^{i},\rvau^{i-1})$,
$i=1,2,\ldots$.
The notation $\mathbb{P}(\rvaa|\rvab,\rvac)$ is to be understood as the conditional distribution of $\rvaa$ given $(\rvab,\rvac)$.
Crucially, this entails the implicit assumption that given $\rvax^{i}$ and $\rvay^{i-1}$, $\rvay(i)$ is independent of every other signal in the system (and likewise for $\mathbb{P}(\rvau(i)|\rvay^{i},\rvau^{i-1})$).
In the representation of Fig.~\ref{fig:diagramas}-(a), this corresponds to assuming $\rvaq\Perp \rvas$ and $(\rvaq,\rvas)\Perp (\rvar,\rvap)$.

\begin{rem}
 The conditioning on the side information $\rvaq$ in both Theorem~\ref{thm:full_D-DPI} and~\cite[Theorem~4.1]{silder11} is motivated by the use of \emph{entropy coded  subtractively dithered quantization} (ECSDQ) in obtaining the upper bound in~\eqref{eq:sdw}.
 For such scenario, the sequences $\rvaq$ and $\rvas $ are identical and correspond to the dither signal, which is independent of $\rvar,\rvap$.
 This satisfies the requirements of~\eqref{eq:full_Dpi} in Theorem~\ref{thm:full_D-DPI} and of~\cite[Theorem~4.1]{silder11}, but not the  assumption that $\rvaq\Perp \rvas$ and $(\rvaq,\rvas)\Perp (\rvar,\rvap)$ implicit in~\cite[Lemma~1]{tanesf18}, which yields~\eqref{eq:tana}.
 In spite of this,~Lemma~1 of~\cite{tanesf18} is used in that paper to prove the lower bound in~\cite[eqn.~8]{tanesf18}, an analogue of~\eqref{eq:sdw} which also considers the use of ECSDQ for the rate term and its upper bound.
\finenunciado
\end{rem}

\subsection{Outline of the Paper}
The remainder of the paper continues with some preliminary definitions and results in Section~\ref{sec:Preliminaries}, followed by the proof of Theorem~\ref{thm:main} in Section~\ref{sec:prfmain}.
Sections~\ref{sec:mad} and~\ref{sec:dad} present additional inequalities relating directed and mutual information, in the former, and two nested directed information expressions, in the latter.
The notions and results associated with in-the-loop channel coding are developed in Section~\ref{sec:itl}.
The main conclusions of this work are presented in Section~\ref{sec:conc}.
Appendix~\ref{sec:fl} has a lemma which is used in proving several of our theorems, while Appendix~\ref{sec:proofs} provides the proofs that are not written right after their corresponding theorems.

An earlier version of this work was made publicly available on arxiv.org~\cite{derost13} and, as such, it has been cited in~\cite{shagra20,tanesf18,silder16,barder20,babgra20}.

\section{Preliminaries}\label{sec:Preliminaries}
\subsection{Mutual Information}\label{subsec:mi}
Let $(\W,\Fsp,P)$ be a probability space, and
$(\Xsp,\Fsp_{\Xsp})$ and $(\Ysp,\Fsp_{\Ysp})$  measurable spaces, and- consider the random variables
$\rvax:\W\to\Xsp$,
$\rvay:\W\to\Ysp$.
Define
$\Msp\eq \Fsp_{\Xsp}\otimes\Fsp_{\Ysp}$, i.e, the $\sigma$-algebra generated by the rectangles 
$\set{A\times B: A\in\Xsp, B\in \Ysp}$.
Consider a probability space 
$(\Xsp\times\Ysp,\Msp,m)$  where $m$ is the (joint) distribution of $(\rvax,\rvay)$  i.e, 
$m=P\circ (\rvax,\rvay)^{-1}$.

Denote the marginal probability distributions of $\rvax$ and $\rvay$ by
$\mu$,
$\nu$, respectively,
where 
\begin{align}
 \mu(A)&=m(A\times \Ysp)
%  \\&=\int 1_{A}(x)dm(x,y),
 ,\fspace A\in\Fsp_{\Xsp}\label{eq:mu_def}\\
 \nu(B)&=m(\Xsp\times B)
%  \\&=\int 1_{B}(y)dm(x,y)
 , \fspace B\in\Fsp_{\Ysp}
\end{align}
Define the product measure $\pi\eq \mu\times\nu$ on $(\Xsp\times\Ysp,\Msp)$.
\begin{defn}
 With the above definitions, the mutual information between $\rvax$ and $\rvay$ is defined as
 \begin{align}
  I(\rvax;\rvay)\eq  \int \log\left( \frac{dm}{d\pi}\right)dm, 
 \end{align} 
\end{defn}
where $\frac{dm}{d\pi}$ is the Radon-Nikodym derivative of $m$ with respect to $\pi$~\cite{yeh---14}.
\begin{lem}[Chain Rule of Mutual Information {\cite[Corollary 7.14]{gray--11}}]
 Suppose $\rvax,\rvay,\rvaz$ are random variables with joint distribution $P_{\rvax\rvay\rvaz}$.
 Suppose also that there exists a product distribution $M_{\rvax\rvay\rvaz}=M_{\rvax}\times M_{\rvay\rvaz}$ such that%
 \footnote{For two probability measures $\mu,\,\nu$ on a common event space $\Uev$ the notation $\mu\ll\nu$ means that $\mu$ is absolutely continuous with respect to $\nu$, i.e., that 
 $\forall\Usp\in\Uev:\mu(\Usp)$}

 $M_{\rvax\rvay\rvaz}\gg P_{\rvax\rvay\rvaz}$.
 (This is true, for example, if $P_{\rvax}\times P_{\rvay\rvaz}\gg P_{\rvax\rvay\rvaz}$.).
 Then
 \begin{align}
  I(\rvax;\rvay,\rvaz)= I(\rvax;\rvay) + I(\rvax;\rvaz|\rvay)
 \end{align}
 The result also holds if $I(\rvax;\rvay,\rvaz)$ is finite.
\end{lem}

The conditional version of the chain rule of mutual information~\cite{yeung-02} (see also~\cite[Corollary~2.5.1]{gray--11}) will be extensively utilized in the proofs of our results:
\begin{align}\label{eq:chainrule_I}
 I(\rvaa,\rvab;\rvac|\rvad) = I(\rvab;\rvac|\rvad) + I(\rvaa;\rvac|\rvab,\rvad).
\end{align}

\subsection{System Description}
We begin by providing a formal description of the systems labeled $\Ssp_{1}\ldots \Ssp_{4}$ in Fig.~\ref{fig:diagramas}-(a). 
Their input-output relationships are given by the possibly-varying deterministic mappings%
\footnote{For notational simplicity, we omit writing their time dependency explicitly.}
\begin{subequations}\label{eq:block_defs}
 \begin{align}
 \rvae(i)&= \Ssp_{1}(\rvau^{i- d_{1}(i)},\rvar^{i}),
\\
\rvax(i)&= \Ssp_{2}(\rvae^{i- d_{2}(i)},\rvap^{i}),
\\
\rvay(i)&= \Ssp_{3}(\rvax^{i- d_{3}(i)},\rvas^{i}),
\\
\rvau(i)&= \Ssp_{4}(\rvay^{i- d_{4}(i)},\rvaq^{i}),
\end{align}
\end{subequations}
where $\rvar,\rvap,\rvas,\rvaq$ are exogenous random signals and the (possibly time-varying) delays $d_{1},d_{2},d_{3},d_{4}\in\set{0,1,\ldots}$ are such that
$$
d_{1}(k)
+
d_{2}(k)+
d_{3}(k)+
d_{4}(k)
\geq  1,\fspace \forall k\in\Nl.
$$
That is, the concatenation of $\Ssp_{1},\ldots,\Ssp_{4}$ has a delay of at least  one sample.
For every $i\in\set{1,\ldots,k}$, $\rvar(i)\in\Rl^{n_{\rvar}(i)}$, i.e., $\rvar(i)$ is a real random vector whose dimension is given by some function $n_{\rvar}:\set{1,\ldots,k}\to \Nl$.
The other sequences ($\rvaq,\rvap,\rvas,\rvax,\rvay,\rvau$) are defined likewise.

\subsection{A Necessary Modification of the Definition of Directed Information}
As stated in~\cite{massey90}, the directed information (as defined in~\eqref{eq:directed_inf_def}) is a more meaningful measure of the flow of information between $\rvax^{k}$ and $\rvay^{k}$ than the conventional mutual information $I(\rvax^{k};\rvay^{k})=\sumfromto{i=1}{k}I(\rvay(i);\rvax^{k}| \rvay^{i-1})$ when there exists causal feedback from $\rvay$ to $\rvax$.
In particular, if $\rvax^{k}$ and $\rvay^{k}$ are discrete-valued sequences, input and output, respectively, of a forward channel, and if there exists \textit{strictly causal}, perfect feedback, so that $\rvax(i)=\rvay(i-1)$  (a scenario utilized in~\cite{massey90} as part of an argument in favor of the directed information), then the mutual information becomes 
\begin{align*}
I(\rvax^{k};\rvay^{k})
&= 
% \Sumfromto{i=1}{k}I(\rvay(i);\rvax^{k}| \rvay^{i-1})
% =
% \Sumfromto{i=1}{k}\left[H(\rvay(i)|\rvay^{i-1}) - H(\rvay(i)|\rvax^{k}, \rvay^{i-1}) \right]
% \\&
% =
H(\rvay^{k}) - H(\rvay^{k}|\rvax^{k})
=
H(\rvay^{k}) - H(\rvay^{k}|\rvay^{k-1})
=
H(\rvay^{k}) - H(\rvay(k)|\rvay^{k-1})
= 
H(\rvay^{k-1}).
\end{align*}
Thus, when strictly causal feedback is present, $I(\rvax^{k};\rvay^{k})$ fails to account for how much information about $\rvax^{k}$ has been conveyed to $\rvay^{k}$ through the forward channel that lies between them.

It is important to note that, in~\cite{massey90} (as well as in many works concerned with communications), the forward channel is instantaneous, i.e., it has no delay.
Therefore, if a feedback channel is utilized, then this feedback channel must have a delay of at least one sample, as in the example above.
However, when studying the system in Fig.~\ref{fig:diagramas}-(a), we may need to evaluate the directed information between signals $\rvax^{k}$ and $\rvay^{k}$ which are, respectively, input and output of a \textit{strictly casual} forward channel (i.e., with a delay of at least one sample), 
whose output is instantaneously fed back to its input.
In such case, if one further assumes perfect feedback and sets $\rvax(i)=\rvay(i)$, then, in the same spirit as before, 
\begin{align*}
I(\rvax^{k}\to \rvay^{k})
&= 
\Sumfromto{i=1}{k}I(\rvay(i);\rvax^{i}| \rvay^{i-1})
=
\Sumfromto{i=1}{k}\left[H(\rvay(i)|\rvay^{i-1}) - H(\rvay(i)|\rvax^{i}, \rvay^{i-1}) \right]
=
H(\rvay^{k}).  
\end{align*}  
As one can see, Massey's definition of directed information ceases to be meaningful if instantaneous feedback is utilized.

It is natural to solve this problem by recalling that, in the latter example, the forward channel had a delay, say $d$, greater than one sample.
Therefore, if we are interested in measuring how much of the information in $\rvay(i)$, not present in $\rvay^{i-1}$, was conveyed from $\rvax^{i}$ through the forward channel, we should look at the mutual information $I(\rvay(i);\rvax^{i-d}|\rvay^{i-1})$, because only the input samples $\rvax^{i-d}$ can have an influence on $\rvay(i)$.
For this reason, 
we introduce
% and in the remainder of this paper, we shall utilize 
the following, modified notion of directed information
\begin{defn}[Directed Information with Forward Delay]
\textit{In this paper, the directed information from $\rvax^{k}$ to $\rvay^{k}$ through a forward channel with a non-negative time varying delay of $d_{xy}(i)$ samples is defined as
\begin{align}
 I(\rvax^{k}\to \rvay^{k}) \eq \Sumfromto{i=1}{k}I(\rvay(i);\rvax^{i-d_{xy}(i)}|\rvay^{i-1}).
\end{align}}
\end{defn}
For a zero-delay forward channel, the latter definition coincides with Massey's~\cite{massey90}.

Likewise, we adapt the definition of causally-conditioned directed information to the definition
\begin{align*}
 I(\rvax^{k}\to \rvay^{k}\parallel\rvae^{k} ) 
 \eq 
 \Sumfromto{i=1}{k}I(\rvay(i);\rvax^{i-d_{xy}(i)}|\rvay^{i-1},\rvae^{i}).
\end{align*}
where, as before, $d_{xy}(i)$ is the delay from $\rvax$ to $\rvay(i)$.

\section{Proof of Theorem~\ref{thm:main}}\label{sec:prfmain}
\begin{figure}[h]
\centering
\input{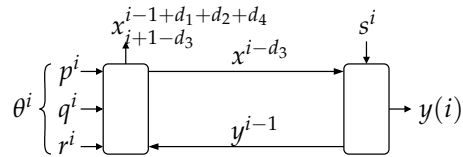}
\caption{Representation of the system of Fig.~\ref{fig:diagramas}-(b) highlighting the dependency between $p$, $q$, $r$, $s$, $x$ and $y$.
The dependency on $i$ of the delays $d_{1}(i),\ldots, d_{4}(i)$ is omitted for clarity.}
\label{fig:Key}
\end{figure}

It is clear from Fig.~\ref{fig:diagramas}-(a) and from~\eqref{eq:block_defs}
that the relationship between $\rvar$, $\rvap$, $\rvaq$, $\rvas$, $\rvax$ and $\rvay$ can be represented by the diagram shown in Fig.~\ref{fig:Key}. 
From this diagram and Lemma~\ref{lem:not_so_obvious} (in Appendix~\ref{sec:fl}) it follows that
if $\rvas$ is independent of $(\rvar,\rvap,\rvaq)$, then the following Markov chain holds:
\begin{align}
 \rvay(i) &
\longleftrightarrow 
(\rvax^{i-d_{3}(i)},\rvay^{i-1})
\longleftrightarrow 
(\rvap^{i},\rvaq^{i},\rvar^{i}).\label{eq:MC1}
\end{align}
Denoting the triad of exogenous signals $\rvap^{k},\rvaq^{k},\rvar^{k}$ by 
\begin{align}
 \theta^{k} \eq (\rvap^{k},\rvaq^{k},\rvar^{k}),
\end{align}
we have the following

\begin{subequations}\label{eq:lanueva}
\begin{align}
 I(\rvax^{k}\to \rvay^{i})
&=
\Sumfromto{i=1}{k}I(\rvay(i);\rvax^{i-d_{3}(i)}|\rvay^{i-1})
\nonumber\\& 
\overset{\eqref{eq:chainrule_I}}{=}
\Sumfromto{i=1}{k}
\left[
I(\theta^{i},\rvax^{i-d_{3}(i)};\rvay(i)|\rvay^{i-1})
-
I(\theta^{i};\rvay(i)|\rvax^{i-d_{3}(i)},\rvay^{i-1})
\right]
\nonumber\\& 
\overset{(a)}{=}
\Sumfromto{i=1}{k}
\left[
I(\theta^{i};\rvay(i)|\rvay^{i-1})
-
I(\theta^{i};\rvay(i)|\rvax^{i-d_{3}(i)},\rvay^{i-1})
\right]\label{eq:solo_directeds}
\\& 
\overset{(b)}{\leq}
\Sumfromto{i=1}{k}
I(\theta^{i};\rvay(i)|\rvay^{i-1})
\overset{(c)}{\leq}
\Sumfromto{i=1}{k}
I(\theta^{k};\rvay(i)|\rvay^{i-1})
\\&=
I(\theta^{k};\rvay^{k}).
\end{align}
\end{subequations}
In the above,
 $(a)$
follows from the fact that, if $\rvay^{i-1}$ is known, then $\rvax^{i-d_{3}(i)}$ is a deterministic function of $\theta^{i}$.
The resulting sums on the right-hand side of~\eqref{eq:solo_directeds} correspond to
$
 I(\rvaq^{k},\rvar^{k},\rvap^{k}\to\rvay^{k})
- 
I(\rvaq^{k},\rvar^{k},\rvap^{k}\to\rvay^{k}\parallel \rvax^{k})
$,
and thereby proving the first part of the theorem, i.e., the equality in~\eqref{eq:main_thm}.
In turn, $(b)$ stems from the non-negativity of mutual informations, turning into equality if $\rvas\Perp(\rvar,\rvap,\rvaq)$, as a direct consequence of the Markov chain in~\eqref{eq:MC1}.
Finally, equality holds in~$(c)$ if $\rvas\Perp(\rvaq,\rvar,\rvap)$, since
$\rvay$ depends causally upon $\theta$.
This shows that equality in~\eqref{eq:main_thm} is achieved if $\rvas\Perp(\rvaq,\rvar,\rvap)$, completing  the proof.
\findemo

\section{Relationships Between Mutual and Directed Informations}\label{sec:mad}

The following result provides an inequality relating $I(\rvax^{k}\to \rvay^{k})$ with the separate flows of information
$I(\rvar^{k} ; \rvay^{k})$ and $I(\rvap^{k},\rvaq^{k}\,;\,\rvay^{k})$.
\begin{thm}\label{thm:from_splitting_more_precise}
\textit{For the system shown in Fig.~\ref{fig:diagramas}-(a), if 
$\rvas\Perp (\rvap,\rvaq,\rvar)$ and 
$\rvar^{k}\Perp (\rvap^{k},\rvaq^{k})$, then 
\begin{align}
I(\rvax^{k}\to \rvay^{k}) 
&
% =
% I(\rvar^{k} ; \rvay^{k})
% +
% I(\rvap^{k},\rvaq^{k};\rvay^{k})
% +
% I(\rvap^{k},\rvaq^{k};\rvar^{k}|\rvay^{k})\label{eq:logro_eq}
% \\&
\geq 
I(\rvar^{k} ; \rvay^{k})+I(\rvap^{k},\rvaq^{k}\,;\,\rvay^{k}). \label{eq:logro}
\end{align}
with equality if and only if the Markov chain $(\rvap^{k},\rvaq^{k})\leftrightarrow\rvay^{k}\leftrightarrow\rvar^{k}$ holds.}
\end{thm}
Theorem~\ref{thm:from_splitting_more_precise} 
shows that, provided $(\rvap,\rvaq,\rvar)\Perp \rvas$, 
$I(\rvax^{k}\to \rvay^{k})$ is lower bounded by the sum of the individual flows from all the subsets in any given partition of $(\rvap^{k},\rvaq^{k},\rvar^{k})$, to $\rvay^{k}$, provided these subsets are mutually independent. 
Indeed,
both theorems~\ref{thm:main} and~\ref{thm:from_splitting_more_precise} can be generalized for any appropriate choice of external and internal signals.
More precisely, let $\Theta$ be the set of all external signals in a feedback system.
Let $\alpha$ and $\beta$ be two internal signals in the loop.
Define $\Theta_{\alpha,\beta}\subset \Theta$ as the set of exogenous signals which are introduced to the loop 
at every subsystem $\Ssp_{i}$ that lies in the path going from $\alpha$ to $\beta$. 
Thus, for any $\rho\in\Theta \setminus \Theta_{\alpha,\beta}$, if $\Theta_{\alpha,\beta} \Perp \Theta\setminus\Theta_{\alpha,\beta}$, we have that~\eqref{eq:main_thm} and~\eqref{eq:logro} become 
\begin{align}
I(\alpha\to\beta)&=
I(\Theta\setminus\set{\Theta_{\alpha,\beta}};\beta),
\\
 I(\alpha\to\beta) - I(\rho;\beta) &\geq I(\Theta\setminus\set{\rho\cup\Theta_{\alpha,\beta}};\beta),
\end{align}
respectively.

To finish this section, we present a stronger, non-asymptotic version of inequality~\eqref{eq:mardah_dir_minus_mutual}:
\begin{thm}\label{thm:three_full_loops}
 \textit{In the system shown in Fig.~\ref{fig:diagramas}-(a), if $(\rvar,\rvap,\rvaq,\rvas)$ are mutually independent, then}
\begin{align}\label{eq:nice}
 I(\rvax^{k}\to\rvay^{k})
&=
I(\rvar^{k};\rvau^{k}) 
+ 
I(\rvap^{k};\rvae^{k})
+ 
I(\rvaq^{k};\rvay^{k}) 
+
I(\rvap^{k};\rvau^{k}|\rvae^{k})
+ 
I(\rvar^{k},\rvap^{k};\rvay^{k}|\rvau^{k}).
\end{align}
\finenunciado
\end{thm}
\begin{rem}\label{rem:mardos}
As anticipated, Theorem~\ref{thm:three_full_loops} can be seen as an extension of~\eqref{eq:mardah_dir_minus_mutual} 
 to the more general setup shown in Fig.~\ref{fig:diagramas}-(a), where the assumptions made 
in~\cite[Lemma~4.1]{mardah05} do not need to hold. 
In particular, letting the decoder $D$ and $\rvap$ in Fig.~\ref{fig:diagramas}-(b) correspond to $\Ssp_{4}$ and $\rvap^{k}$ in Fig.~\ref{fig:diagramas}-(a), respectively, we see that inequality~\eqref{eq:mardah_dir_minus_mutual} holds even if the channel $f$ has memory or $D$ and $E$ have independent initial states, or 
if the internal state of $D$ is not observable~\cite{gogrsa01}.
\finenunciado
\end{rem}

\begin{figure}[htbp]
 \centering
 \input{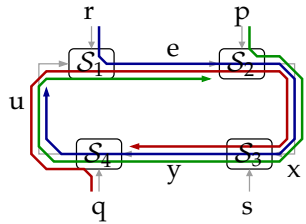}
 \caption{A representation of the three first information flows on the right-hand-side of~\eqref{eq:nice}.}
 \label{fig:flujos2}
\end{figure}

Theorem~\ref{thm:three_full_loops} also admits an interpretation in terms of information flows.
This can be appreciated in the diagram shown in Fig.~\ref{fig:flujos2}, which 
depicts the individual full-turn flows (around the entire feedback loop) stemming from $\rvaq$, $\rvar$ and $\rvap$.
Theorem~\ref{thm:three_full_loops} states that the sum of these individual flows is a lower bound for the 
directed information from $\rvax$ to $\rvay$, provided $\rvaq,\rvar,\rvap,\rvas$ are independent.

\section{Relationships Between Nested Directed Informations}\label{sec:dad}
This section presents three closed-loop versions of the data-processing inequality \textit{relating two directed informations}, both between pairs of signals \textit{internal} to the loop.
As already mentioned in Section~\ref{sec:intro},
to the best of our knowledge, the first inequality of this type to appear in the literature is the one in
Theorem~4.1 in~\cite{silder11} (see~\eqref{eq:silder11}).
Recall that the latter result stated that 
$I(\rvax^{k}\to\rvay^{k}\parallel \rvaq^{k})\geq I(\rvax^{k}\to\rvau^{k})$,
requiring
$\Ssp_{4}$ to be such that $\rvay^{i}$ is a deterministic function of $(\rvau^{i},\rvaq^{i})$ and that $\rvaq\Perp(\rvar,\rvap)$.
The following result presents another inequality which also relates two nested directed informations, namely, 
$
I(\rvax^{k}\to \rvay^{k})$ and $I(\rvae^{k}\to \rvay^{k})
$,  
but requiring only that $\rvas\Perp (\rvaq,\rvar,\rvap)$. 
\begin{thm}\label{thm:DPI_dir_dir}
\textit{For the closed-loop system in Fig.~\ref{fig:diagramas}-(b), if 
$(\rvaq,\rvar,\rvap)\Perp \rvas$, then
 \begin{align}
  I(\rvax^{k}\to \rvay^{k}) &\geq I(\rvae^{k}\to \rvay^{k}).
 \end{align}}
 \finenunciado
\end{thm}
Notice that Theorem~\ref{thm:DPI_dir_dir} does not require $\rvap$ to be independent of $\rvar$ or $\rvaq$.
This may seem counter-intuitive upon noting that $\rvap$ enters the loop between the link from $\rvae$ to $\rvax$. 

The following theorem is an identity between two directed informations involving only internal signals.
It can also be seen as a  complement to Theorem~\ref{thm:DPI_dir_dir}, since it can be directly applied to establish the relationship
between $ I(\rvae^{k}\to \rvay^{k})$
and $I(\rvae^{k}\to \rvau^{k})$.
%START DPI DIRECTED SECOND PART

\begin{thm}\label{thm:finally}
\textit{For the system shown in Fig.~\ref{fig:diagramas}-(a), 
if $\rvas\Perp (\rvaq,\rvar,\rvap)$, then
\begin{align}\label{eq:finally0}
 I(\rvax^{k}\to \rvay^{k})
\geq 
I(\rvax^{k}\to \rvau^{k})
+
I(\rvaq^{k}\,;\, \rvay^{k})
+
I(\rvar^{k},\rvap^{k}; \rvay^{k}| \rvau^{k})
+ I(\rvaq^{k};\rvar^{k}|\rvau^{k},\rvay^{k}).
\end{align}
with equality if, in addition, $\rvaq\Perp(\rvar,\rvap)$. 
In the latter case, it holds that
\begin{align}\label{eq:finally}
I(\rvax^{k}\to \rvay^{k})
=
I(\rvax^{k}\to \rvau^{k})
+
I(\rvaq^{k}\,;\, \rvay^{k})
+
I(\rvar^{k},\rvap^{k}; \rvay^{k}| \rvau^{k}).
\end{align}
}\finenunciado
\end{thm}
Notice that, by requiring additional independence conditions upon the exogenous signals (specifically, $\rvaq\Perp\rvas$), Theorem~\ref{thm:finally} (and, in particular,~\eqref{eq:finally})  yields 
\begin{align}\label{eq:la_algo_mejor}
I(\rvax^{k}\to \rvay^{k})
\geq
I(\rvax^{k}\to \rvau^{k}),
\end{align}
which strengthens 
the inequality in~\cite[Theorem~4.1]{silder11} (stated above in~\eqref{eq:silder11}).
More precisely,~\eqref{eq:la_algo_mejor} does not require conditioning one of the directed informations and holds irrespective of the invertibility of the mappings in the loop.

\section{Giving Operational Meaning to the Directed Information: In-the-loop Channel Coding}\label{sec:itl}
In this section we introduce the notions of in-the-loop transmission rate and capacity and show that they are related by the directed information rate across the channel in the same feedback loop.
This provides another example to illustrate the applicability of theorems~\ref{thm:main} and~\ref{thm:full_D-DPI} and also provides further operational meaning to the directed information rate.

\begin{figure}[htbp]
\centering
\begin{tikzpicture}[node distance=9mm]
 \def\la{3.5mm}
\node (b1) [bloq] {$\Esp$};
\node (b2) [bloq, right =of b1] {$\Csp$};
\node (b3) [bloq, right = of b2, xshift=7mm] {$\Dsp$};
\node (b4) [bloq, below =of b2, yshift=3mm] {$\Ssp$};

\path (b1) edge[->] node[pos=0.5, above] {$\rvax$} (b2);
\path (b2) edge[->] node[pos=0.5, above] {$\rvay$} (b3);
\draw[->] (b2.east) --++ (8mm,0) |-($(b4.east) $) node[pos=.75,below] {$\rvay$};
\draw[->] (b4.west) --++(-2.1,0) |-(b1.west) node[ pos=.75, above] {$\rvaw$};

\draw[->] (b3.east) --++(\la,0)  node[ right] {$\rvav$};

\draw[<-] (b1.north)--++(0,\la) node[above] {$\rvar$};
\draw[<-] (b2.north)--++(0,\la) node[above] {$\rvas$};
\draw[<-] (b3.north)--++(0,\la) node[above] {$\rvap$} coordinate (ps);
\draw[<-] (b4.south)--++(0,-\la) node[below] {$\rvaq$} coordinate (pq);
\end{tikzpicture}\caption{A communication feedback  system in which the messages and the channel output are within the loop.  } 
\label{fig:diagrama}
\end{figure}
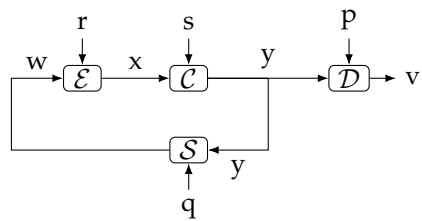

Consider the scheme shown in Fig.~\ref{fig:diagrama}, and suppose $\Csp$ is a noisy communication channel. 
Let $\Esp$ and $\Dsp$ be channel encoder and decoder, respectively, with $\rvar$ and $\rvap$ being side information sequences causally and independently available to each of them such that $(\rvar,\rvap)\Perp(\rvas,\rvaq)$.
This means that, for $k=1,2,\ldots,n$,
\begin{align}
 (\rvap_{1}^{n},\rvar_{k+1}^{n})&\leftrightarrow \rvar_{1}^{k} \leftrightarrow (\rvaw_{1}^{k+1},\rvax_{1}^{k},\rvay_{0}^{k}).\\
 \rvap_{k+1}^{n}&\leftrightarrow \rvap_{1}^{k} \leftrightarrow \rvar_{1}^{k}\\
 \rvap_{k+1}^{n}&\leftrightarrow \rvap_{1}^{k} \leftrightarrow (\rvaw_{1}^{k+1},\rvax_{1}^{k},\rvay_{0}^{k}).\label{eq:mcppall}
\end{align}
A crucial aspect of this scenario is the fact that the messages $\rvaw_{1}^{n}, \rvaw_{n+1}^{2n},\ldots$ to be encoded are contained in the sequence $\rvaw$, a signal \textbf{internal to the loop}; they can be regarded as a corrupted version of the decoded messages, which comprise the sequence $\rvav$.
This is a key difference with respect to the available literature on feedback capacity, where, to the best of the authors' knowledge, the messages are  exogenous and the feedback signal only helps in the  encoding task.%
\footnote{Exceptions can be found in some papers on networked control which consider in-the-loop channel coding, such as, e.~g.,~\cite{sahmit06,khigar19}.}
In Fig.~\ref{fig:diagrama}, the latter standard scenario corresponds to encoding the sequence $\rvar$.

The fact that the messages to be  encoded bear information from the decoded message symbols requires one to redefine the notion of information transmission rate commonly used in the standard scenario.
To see this, let $\rvaw(k)\in\Wsp, \ k=1,2,\ldots$, for some finite alphabet $\Wsp$ of cardinality $\abs{\Wsp}$, and notice that the transmission rate definitions  $\log(\abs{\Wsp})$ and $H(\rvaw_{1}^{n})/n$ are unsatisfactory if $\rvaw(k)=\rvay(k-1),\ k=1,2,\ldots$, i.e., if the messages to be transmitted are already available at the decoder (more generally, if there is no randomness in the feedback path).
This suggests that a suitable notion of transmission rate for this scenario should exclude information that is \textbf{already known by the receiver}.

In view of the above, we propose the following notion of transmission rate for the case in which the messages to be transmitted are in the loop:
  \begin{defn}
For the system described in Fig.~\ref{fig:diagramas}, the in-the-loop (ITL) transmission rate is defined as
 \begin{align}
  R_{\text{ITL}}^{n}\eq \frac{1}{n}\Sumfromto{k=1}{n}H(\rvaw(k)|\rvaw_{1}^{k-1},\rvay_{0}^{k-1},\rvap^{k})
 \end{align}
 \finenunciado
 \end{defn}
 The meaning of the ITL transmission rate is further elucidated by  considering the  following scenarios:
 \begin{enumerate}
\item If the feedback channel is deterministic, then $\rvaw(k)$ is a deterministic function of $\rvay_{0}^{k-1}$ and thus  $R_{\text{ITL}}^{n}=0$, as desired.
\item If the (forward) communication channel is noiseless, then at each time $k-1$, we have $\rvay_{1}^{k-1}=\rvaw_{1}^{k-1}$.
Therefore $R_{\text{ITL}}^{n}=H(\rvaw_{1}^{n}|\rvay_{0},\rvap_{1}^{n})/n$. 
Again, if the feedback channel is deterministic, the ITL transmission rate is zero.
\item In the absence of feedback, 
$R_{\text{ITL}}^{n}= \frac{1}{n}H(\rvaw_{1}^{n})$, recovering the notion of transmission rate of the case in which the messages are exogenous to the loop.
 \end{enumerate}
Thus, $R_{\text{ITL}}^{n}$ can be interpreted  as \textbf{the sum of the information the encoder attempts to transmit at each sample time that is novel for both the transmitter and    the receiver. }

\begin{thm}\label{thm:rnlb}
Consider the setup depicted in Fig.~\ref{fig:diagrama}, where $\Esp$ and $\Dsp$ are channel encoder and decoder, respectively, and $\Csp$ is the communication channel.
Suppose the message and side-information samples $\rvaw(k)\in\Wsp,\ \rvar(k)\in\Rsp ,\ k=1,2,\ldots$, respectively where $\Wsp$ and $\Rsp$ are  finite alphabets.
Define the binary random variable $\rvae_{n}$ to equal $1$ if $\rvav_{1}^{n}\neq\rvaw_{1}^{n}$ and $0$ otherwise. 
Then, for every $n\in\Nl$,
\begin{align}
 R_{\text{ITL}}^{n}\ge I(\rvaw_{1}^{n}\to\rvay_{0}^{n}\Vert\rvap_{1}^{n}) ,
\end{align}
with equality if and only if $H(\rvaw_{1}^{n}|\rvay_{0}^{n},\rvap_{1}^{n})=0$.
Moreover,
\begin{align}\label{eq:perr}
 \Pr\set{\rvae_{n}=1}
 &=
 \frac{ 
 \rin
 -
 \frac{1}{n}I(\rvaw_{1}^{n}\to\rvay_{0}^{n}\Vert\rvap_{1}^{n})
 -
\frac{1}{n}H(\rvae_{n}|\rvay_{0}^{n},\rvap_{1}^{n})
}{
   \frac{1}{n}H(\rvaw_{1}^{n}|\rvay_{0}^{n},\rvap_{1}^{n},\rvae_{n}=1)
}
 \\&
 \ge \frac{R_{\text{ITL}}^{n}-\frac{1}{n}
 I(\rvaw_{1}^{n}\to\rvay_{0}^{n}\Vert\rvap_{1}^{n})-1/n}{\log_2(\abs{\Wsp} )}
\end{align}
\end{thm}

\begin{proof}
Recall that
\begin{align}
I(\rvaw_{1}^{n}\to\rvay_{0}^{n}\Vert\rvap_{1}^{n}) 
&=
\Sumfromto{k=1}{n}I(\rvaw_{1}^{k};\rvay(k)|\rvay_{0}^{k-1},\rvap_{1}^{k})
=
\Sumfromto{k=1}{n}H(\rvaw_{1}^{k} |\rvay_{0}^{k-1},\rvap_{1}^{k})
-\Sumfromto{k=1}{n}H(\rvaw_{1}^{k} |\rvay_{0}^{k},\rvap_{1}^{k})\label{eq:ito}
\end{align}

On the other hand,
\begin{align}
 nR_{\text{ITL}}^{n}&=
 \Sumfromto{k=1}{n}H(\rvaw(k)|\rvaw_{1}^{k-1},\rvay_{0}^{k-1},\rvap_{1}^{k})
 \overset{\text{(cr)}}{=} \Sumfromto{k=1}{n}H(\rvaw_{1}^{k}|\rvay_{0}^{k-1},\rvap_{1}^{k})
 -
 \Sumfromto{k=2}{n}H(\rvaw_{1}^{k-1}|\rvay_{0}^{k-1},\rvap_{1}^{k})
 \\&
 \overset{\eqref{eq:ito}}{=}
 \Sumfromto{k=1}{n}H(\rvaw_{1}^{k} |\rvay_{0}^{k},\rvap_{1}^{k})
 -
 \Sumfromto{k=2}{n}H(\rvaw_{1}^{k-1}|\rvay_{0}^{k-1},\rvap_{1}^{k})
 +
I(\rvaw_{1}^{n}\to\rvay_{0}^{n}\Vert\rvap_{1}^{n}) 
\\&
 \overset{\eqref{eq:mcppall}}{=}
 H(\rvaw_{1}^{n}|\rvay_{0}^{n},\rvap_{1}^{n})
 +
I(\rvaw_{1}^{n}\to\rvay_{0}^{n}\Vert\rvap_{1}^{n}) \label{eq:before},
\end{align}
where the equality (cr) follows from the chain rule of entropy.
This proves the first part of the theorem.

Let us now re-derive the first steps leading to Fano's inequality, to include the side-information $\rvap_{1}^{n}$ and to verify that it is not affected by the fact that $\rvaw$ and $\rvay$ are within the loop.
\begin{align}
 H(\rvaw_{1}^{n}|\rvay_{0}^{n},\rvap_{1}^{n})
 &
 \overset{\text{(cr)}}{=}
 H(\rvaw_{1}^{n},\rvae_{n}|\rvay_{0}^{n},\rvap_{1}^{n})
-
 H(\rvae_{n}|\rvay_{0}^{n},\rvap_{1}^{n},\rvaw_{1}^n)
 \\&\overset{(a)}{=}
 H(\rvae_{n}|\rvay_{0}^{n},\rvap_{1}^{n})
 +
 H(\rvaw_{1}^{n}|\rvay_{0}^{n},\rvap_{1}^{n},\rvae_{n})
 \\&
 \overset{(b)}{=}
 H(\rvae_{n}|\rvay_{0}^{n},\rvap_{1}^{n})
 +
  H(\rvaw_{1}^{n}|\rvay_{0}^{n},\rvap_{1}^{n},\rvae_{n}=1)\Pr\set{\rvae_{n}=1},
\end{align}
where $(a)$ holds because
$ H(\rvae_{n}|\rvay_{0}^{n},\rvap_{1}^{n},\rvaw_{1}^n)=0$ and from the chain rule, while $(b)$ is because
$H(\rvaw_{1}^{n}|\rvay_{0}^{n},\rvap_{1}^{n},\rvae_{n}=0)=0$.

Substituting this into~\eqref{eq:before},
\begin{align}
nR_{\text{ITL}}^{n}&=  
H(\rvae_{n}|\rvay_{0}^{n},\rvap_{1}^{n})
 +
   H(\rvaw_{1}^{n}|\rvay_{0}^{n},\rvap_{1}^{n},\rvae_{n}=1)\Pr\set{\rvae_{n}=1}
 +
I(\rvaw_{1}^{n}\to\rvay_{0}^{n}\Vert\rvap_{1}^{n}).
\end{align}
Noting that $H(\rvae_{n}|\rvay_{0}^{n},\rvap_{1}^{n})\le 1$ and
$H(\rvaw_{1}^{n}|\rvay_{0}^{n},\rvap_{1}^{n},\rvae_{n}=1)\le n \log(\abs{\Wsp})$ leads directly to~\eqref{eq:perr}, comparing the proof.
\end{proof}

Theorem~\ref{thm:rnlb} allows one to draw an additional interpretation of the ITL transmission rate.
We extend first the identity of~\cite{masmas05} to include causal conditioning by $\rvap_{1}^{n}$:
\begin{align}\label{eq:split}
I(\rvaw_{1}^{n};\rvay_{0}^{n},\rvap_{1}^{n})
 &=
  I(\rvaw_{1}^{n};\rvay_{n}|\rvay_{0}^{n-1},\rvap_{1}^{n})
+ 
 I(\rvaw_{n};\rvay_{0}^{n-1}|\rvaw_{1}^{n-1},\rvap_{1}^{n})
 +
 I(\rvaw_{1}^{n-1};\rvay_{0}^{n-1},\rvap_{1}^{n-1})
 \\&
 =
 \Sumfromto{k=1}{n}
 I(\rvaw_{1}^{k};\rvay(k)|\rvay_{0}^{k-1},\rvap_{1}^{k})
+ 
 \Sumfromto{k=1}{n}
 I(\rvaw(k);\rvay_{0}^{k-1}|\rvaw_{1}^{k-1},\rvap_{1}^{k})
 =
I(\rvaw_{1}^{n}\to\rvay_{0}^{n}\Vert\rvap_{1}^{n})
+
I(\rvay_{0}^{n-1}\to\rvaw_{1}^{n}\Vert\rvap_{1}^{n}),
  \end{align}
  where 
  \begin{align}
I(\rvay_{0}^{n-1}\to\rvaw_{1}^{n}\Vert\rvap_{1}^{n})  
\eq  
   \Sumfromto{k=1}{n}
  I(\rvaw(k);\rvay_{0}^{k-1}|\rvaw_{1}^{k-1},\rvap_{1}^{k}).
        \end{align}
        It readily follows from~\eqref{eq:split} that
        \begin{align}
         H(\rvaw_{1}^{n})
         -I(\rvay_{0}^{n-1}\to\rvaw_{1}^{n}\Vert\rvap_{1}^{n})
=
I(\rvaw_{1}^{n}\to\rvay_{0}^{n}\Vert\rvap_{1}^{n})
+
         H(\rvaw_{1}^{n}|\rvay_{0}^{n},\rvap_{1}^{n})
         \overset{\eqref{eq:before}}{=}
         n\rin.
        \end{align}
Thus, the ITL transmission rate corresponds to the entropy rate of the messages having extracted from it the information flowing from the decoder input to the messages.

The main result of this section is the following theorem, which asserts that the supremum of achievable ITL transmission rates is upper bounded by the directed information across the communication channel.
\begin{thm}\label{thm:itlrcap}
Consider the setup depicted in Fig.~\ref{fig:diagrama}, where $\Esp$ and $\Dsp$ are channel encoder and decoder, respectively, and $\Csp$ is the communication channel.
Then the supremum of achievable ITL transmission rates is upper bounded by the supremum of the directed information rate from $\rvax$ to $\rvay$ causally conditioned by $\rvap_{1}^{n}$.
\end{thm}
\begin{proof}
 The result follows directly from Theorem~\ref{thm:rnlb} and from Theorem~\ref{thm:full_D-DPI}.
\end{proof}

Thus, the supremum of $\lim_{n\to\infty}I(\rvax_{1}^{n}\to\rvay_{1}^{n}\Vert\rvap_{1}^{n})$ is an outer bound to the capacity region of ITL transmission rates.

In the following example, this bound is reachable.

\begin{exple}\label{ex:ritl}
Consider the case in which the forward channel $\Csp$ in Fig.~\ref{fig:diagrama} is transparent, i.e., $\rvay(k)=\rvax(k)$ for $k=0,1,\ldots$, as shown in Fig.~\ref{fig:diagramf}.
 Let $\rvay(k)\in\set{0,1,2,3}$, $k=0,1,\ldots$.
 Let $\rvaq(0)=1$ (deterministically) and $\rvaq(1),\rvaq(2),\ldots$ be binary and i.i.d. with $\Pr\set{\rvaq(k)=1}=\alpha=0.9$.
 The feedback channel $\Ssp$ is defined by the following recursion
 \begin{align}
  \rvaw(k)&=
  \begin{cases}
  \rvaq(k)&,\ \text{ if } \rvaq(k-1)=(\rvay(k-1)\mod 2)
  \\
   (\rvay(k-1)\mod{2})&\ \text{, if } \rvaq(k-1)\neq(\rvay(k-1)\mod 2)
  \end{cases}
  \fspace,\ k=1,2,\ldots
 \end{align}
Thus, $\Ssp$ outputs a new sample of $\rvaq$ iff the previous sample of $\rvaq$ is matched by the previous sample $\mod{2}$ of $\rvay$.
Otherwise, it lets $\rvay(k-1)\mod{2}$ pass through.

\begin{figure}[htbp]
\centering
% u--to v
% e--to w
\begin{tikzpicture}[node distance=9mm]
 \def\la{3.5mm}
\node (b1) [bloq] {$\Esp_{1}$};
\node (b3) [bloq, right = of b1, xshift=27mm] {$\Dsp_{1}$};
\node (b4) [bloq, below =of b1, yshift=3mm,xshift=10mm] {$\Ssp$};

\path (b1) edge[->] node[pos=0.5, above] {$\rvay$} (b3);
\draw[->] (b1.east) --++ (28mm,0) coordinate (c) |-($(b4.east) $) node[pos=.75,below] {$\rvay$};
\draw[->] (b4.west) --++(-2.1,0) |-(b1.west) node[ pos=.75, above] {$\rvaw$};

\draw[->] (b3.east) --++(\la,0)  node[ right] {$\rvav$} coordinate (pv);

\draw[<-] (b1.north)  --++(0,\la) node[above] {$\rvar$} coordinate (a);
\draw[<-] (b4.south)--++(0,-\la) node[below] {$\rvaq$} coordinate (pq);

\def\gap{1.5mm}
\coordinate (h) at (\gap,0);
\coordinate (v) at (0,\gap);
\coordinate (b) at ($0.5*(b1.east)+0.5*(b1.west)$);
\coordinate (cc) at (b3.east);
\coordinate (f) at ($(b4.west)-(2.1,0)$);
\tikzmath{
    coordinate \c, \d, \f;
    \c=(b4.east);
    \d=(c);
    \f=(f);
}
\coordinate (d) at(\dx,\cy);
\coordinate (e) at ($0.5*(b4.west)+0.5*(b4.east)$);
\coordinate (g) at (\fx,0);

\draw[->,red,opacity=0.5,
line width=0.8mm, 
% ultra thick,
% arrows={->Latex[width'=2pt]}
]
        (a)
    -- ($(b)+(v)$)
    -- ($(b)+(h)$)
    -- ($(c) -(h)$) coordinate (ch)
    -- ($(c) -(v)$)
    -- ($(d) +(v)$)
    -- ($(d) -(h)$)
    -- (e)
    ;
    \draw[->,red,opacity=0.5,very thick]
    (e)
        -- ($(b4.west)-(1,0)$)
;
    
    \draw[->,red,opacity=0.5,very thick] 
        (ch)
    -- ($(pv)+(0,0)$)
    ;
    \def\ss{.07}
\coordinate (sh) at (\ss,0);
\coordinate (sv) at (0,\ss);
\draw[->,green,opacity=0.6,line width=.8mm]%,arrows={-latex[width=2pt]}]
        (pq)
    -- ($(e)-(v)-(sv)$)
    -- ($(e)-(h)-(sv)$) coordinate (qb)
;
\draw[->,green,opacity=0.7, thick] 
        (qb)
    -- ($(f)+(h)-(sv)$)
    -- ($(f)+(v)-(sh)$)
    -- ($(g)-(v)-(sh)$)
    -- ($(g)+(h)+(sv)$)
    -- ($(pv)+(sv)$)
    ;
    \node at ($(pq)-(0,1)$) {(a)};
\end{tikzpicture}
\hspace{5mm}
\begin{tikzpicture}[node distance=9mm]
 \def\la{3.5mm}
\node (b1) [bloq] {$\Esp_{2}$};
\node (b3) [bloq, right = of b1, xshift=27mm] {$\Dsp_{2}$};
\node (b4) [bloq, below =of b1, yshift=3mm,xshift=10mm] {$\Ssp$};

\path (b1) edge[->] node[pos=0.5, above] {$\rvay$} (b3);
\draw[->] (b1.east) --++ (28mm,0) coordinate (c) |-($(b4.east) $) node[pos=.75,below] {$\rvay$};
\draw[->] (b4.west) --++(-2.1,0) |-(b1.west) node[ pos=.75, above] {$\rvaw$};

\draw[->] (b3.east) --++(\la,0)  node[ right] {$\rvav$} coordinate (pv);

% \draw[<-] (b1.north)  --++(0,\la) node[above] {$\rvar$} coordinate (a);
\draw[<-] (b4.south)--++(0,-\la) node[below] {$\rvaq$} coordinate (pq);

\def\gap{1.5mm}
\coordinate (h) at (\gap,0);
\coordinate (v) at (0,\gap);
\coordinate (b) at ($0.5*(b1.east)+0.5*(b1.west)$);
\coordinate (cc) at (b3.east);
\coordinate (f) at ($(b4.west)-(2.1,0)$);
\tikzmath{
    coordinate \c, \d, \f;
    \c=(b4.east);
    \d=(c);
    \f=(f);
}
\coordinate (d) at(\dx,\cy);
\coordinate (e) at ($0.5*(b4.west)+0.5*(b4.east)$);
\coordinate (g) at (\fx,0);

    \def\ss{.07}
\coordinate (sh) at (\ss,0);
\coordinate (sv) at (0,\ss);
\draw[->,green,opacity=0.6,line width=.8mm]%,arrows={-latex[width=2pt]}]
        (pq)
    -- ($(e)-(v)-(sv)$)
    -- ($(e)-(h)-(sv)$) coordinate (qb)
;
\draw[->,green,opacity=0.7, line width=.8mm] 
        (qb)
    -- ($(f)+(h)-(sv)$)
    -- ($(f)+(v)-(sh)$)
    -- ($(g)-(v)-(sh)$)
    -- ($(g)+(h)+(sv)$)
    -- ($(pv)+(sv)$)
    ;
    \node at ($(pq)-(0,1)$) {(b)};
\end{tikzpicture}

\caption{The feedback communication system considered in Example~\ref{ex:ritl}.
In (a), encoder/decoder pair~1 yields a large mutual information between $\rvaw$ and $\rvav$ by adding a backward information flow (in red) to the forward information flow (in green). The latter is only part of thee entropy rate of $\rvaq$.
In (b), encoder/decoder pair~2 yields a smaller mutual information between $\rvaw$ and $\rvav$, but it corresponds to the greatest possible forward information flow, which coincides with the entropy rate of $\rvaq$.
Thus, it is capacity achieving with respect to the ITL transmission rate.} 
\label{fig:diagramf}
\end{figure}
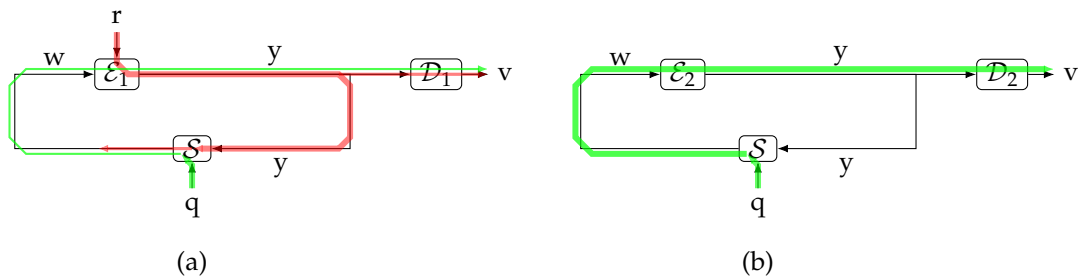

Consider first the following encoder-decoder pair, designed with the aim of achieving zero-error communication while maximizing $H(\rvaw_{1}^{n})/n=I(\rvaw_{1}^{n};\rvav_{1}^{n})$.

Encoder $\Esp_{1}$:
Let the side-information sequence
$\rvar$ be binary i.i.d. and independent of $\rvaq$, with $\Pr\set{\rvar(k)=1}=\beta$, and
\begin{align}
\rvay(0)&=\rvar(0)\\
 \rvay(k)&=
 \begin{cases}
  \rvar(k)& \text{, if } \rvaw(k)=(\rvay(k-1)\mod{2})
  \\
  \rvar(k)+2 & \text{, if } \rvaw(k)\neq(\rvay(k-1)\mod{2})
 \end{cases}
 \fspace,\ k=1,2,\ldots
\end{align}
Decoder $\Dsp_{1}$:
\begin{align}
 \rvav(k)&=
 \begin{cases}
\rvay(k-1)\mod{2} &\text{, if  } \rvay(k)\leq1
\\
(\rvay(k-1)\mod{2})\oplus 1 &\text{, if  } \rvay(k)>1
\end{cases}
\end{align}
where $\oplus$ is the exclusive-or binary operator.
With this choice, $\rvav(k)=\rvaw(k)$ for $k=1,2,\ldots$.
In addition,
\begin{align}\label{eq:wencone}
\rvaw(k) &=
\begin{cases}
 \rvaq(k) &\text{, if } \rvar(k-1)= \rvaq(k-1)
\\
 \rvar(k-1)&\text{, if } \rvar(k-1)\neq \rvaq(k-1)
\end{cases}\fspace,\ k=1,2,\ldots
\end{align}
Therefore,
\begin{align}
 \Pr\set{\rvaw(1)=1}=\alpha\beta.
\end{align}
and, for $k\geq2$,
\begin{align}
 \Pr&\set{\rvaw(k)=1}
 =
(\alpha^{2}+(1-\alpha)^{2})\beta+\alpha(1-\alpha)
\end{align}
Thus, and since $\alpha=0.9$, the entropy of each $\rvaw(k)$ is maximized by 
$\beta
%=0.41/0.811
=0.5055$.
However, encoder $\Esp_{1}$ makes the samples of $\rvaw$  interdependent, so finding the value of $\beta$ that maximizes 
$H(\rvaw_{1}^{n})/n$ (and thus $I(\rvaw_{1}^{n};\rvav_{1}^{n})$ as well) is more involved, and that value  does not need to be the same.
We have found numerically that (for $n=22$) the maximum of 
$H(\rvax_{1}^{n})/n=I(\rvaw_{1}^{n};\rvav_{1}^{n})/n$ is (approximately) 0.9941 [bits/sample], attained with $\beta=0.503$, very close to the $\beta$ which maximizes $H(\rvaw_{1}^{n})/n$.

For later comparison, we also calculate the value of  $\rin$ yielded by this choice of encoder:
\begin{align}
 \rin
 \overset{(a)}{=}
 I(\rvaw_{1}^{n}\to\rvay_{1}^{n})
 \overset{\text{Thm. \ref{thm:main}}}{=}
I(\rvaq_{0}^{n};\rvay_{0}^{n})
 =
 \Sumfromto{k=0}{n}( H(\rvaq(k)|\rvaq_{0}^{k-1})-H(\rvaq(k)|\rvaq_{0}^{k-1},\rvay_{0}^{n})),
\end{align}
where $(a)$ holds from Theorem~\ref{thm:rnlb} because $H(\rvaw_{1}^{n}|\rvay_{0}^{n})=0$.
Defining the binary random variables
$\rvaa(k)\eq1$ when $(\rvay(k)\mod{2})=\rvaq(k)$ and $0$ otherwise, we get
\begin{align}
 H(\rvaq(k)|\rvaq_{0}^{k-1},\rvay_{0}^{n})
 &=
 H(\rvaq(k)|\rvaq_{0}^{k-1},\rvay_{0}^{n},\rvaa(k-1))
 \\&
 =
 H(\rvaq(k)|\rvaq_{0}^{k-1},\rvay_{0}^{n},\rvaa(k-1)=0) \Pr\set{\rvaa(k-1)=0}
+
 H(\rvaq(k)|\rvaq_{0}^{k-1},\rvay_{0}^{n},\rvaa(k-1)=1) \Pr\set{\rvaa(k-1)=1}\nonumber
 \\&
 \overset{\eqref{eq:wencone}}{=}
 H(\rvaq(k))\Pr\set{\rvaa(k-1)=0}
+
 0\cdot \Pr\set{\rvaa(k-1)=1}
\end{align}
Thus
\begin{align}
 \rin&=H(\rvaq(k))(1-\Pr\set{\rvaa(k-1)=0})
 =
 H(\rvaq(k))(\alpha(1-\beta)+(1-\alpha)\beta)
 \\&=
 0.469\times 
 0.4976
 =0.2334\fspace\text{[bits/sample]},
\end{align}
using $\beta=0.503$.

The second encoder/decoder pair is set to maximize $R_{\text{ITL}}^{n}$, and is defined as follows:

Encoder $\Esp_{2}$:
\begin{align}
 \rvay(k)=
 \begin{cases}
 1              & \text{, if }            k=0\\
 \rvaw(k)  &\text{, if }k\geq1
\end{cases}
\end{align}
Thus, zero-error communication is trivially attained with the simple decoding rule:

Decoder $\Dsp_{2}$:
\begin{align}
\rvav(k)=\rvay(k) , \fspace k\geq 1.
\end{align}
Also, encoder $\Esp_{2}$ yields 
$\rvaw(k)=\rvaq(k)$, for $k\geq1$ .
Therefore, 
\begin{align}
 \frac{1}{n}I(\rvaw_{1}^{n}\to\rvay_{0}^{n})
 \overset{\text{Thm.~\ref{thm:rnlb}}}{=} 
 R_{\text{ITL}}^{n}&
 =\frac{1}{n}H(\rvaq_{1}^{n})
 =0.469\fspace\text{[bits/sample]}
\end{align}
As expected, encoder $\Esp_{2}$ yields a higher $R_{\text{ITL}}^{n}$ than encoder $\Esp_{1}$.
More significant is the fact that\textbf{ encoder/decoder pair~2 achieves the in-the-loop  capacity for this channel,} since
\begin{align}
  \frac{1}{n}I(\rvaw_{1}^{n}\to\rvay_{0}^{n})
\overset{(a)}{=}
I(\rvaq_{1}^{n};\rvay_{0}^{n})
\leq H(\rvaq_{1}^{n})
\end{align}
\end{exple}

The previous example illustrates an important fact, closely related with the motivation behind the definition of $\rin$:
maximizing the mutual information between the messages to be transmitted and the decoded messages (a leitmotif in traditional channel coding, wherein messages are generated outside the loop) is not suitable when messages are in the loop.

Indeed,~\eqref{eq:split} provides a mathematically precise meaning to the above observation.
It reveals why maximizing $I(\rvaw_{1}^{n};\rvay_{0}^{n},\rvap_{1}^{n})$ does not necessarily mean maximizing $I(\rvaw_{1}^{n}\to\rvay_{0}^{n}\Vert\rvap_{1}^{n})$, since the former is the sum of backward and forward information flows (represented in green and red in Fig.~\ref{fig:diagramf}, respectively).

Finally, theorems~\ref{thm:rnlb} and~\ref{thm:itlrcap} imply that in the design of any encoder for in-the-loop messages, aiming to yield the joint probability distribution of channel input and output sequences that maximizes the directed information is of practical importance:
it is  necessary for achieving the highest ``useful'' transmission rate while minimizing the probability of error.

\section{Conclusions}\label{sec:conc}
The widely used data processing inequality does not hold for systems with feedback. 
In this work, we provided a very general \emph{directed information} data processing inequality that is applicable to feedback systems. 
A key insight to be gained from this new inequality is that, for nested pairs of sequences, the further apart the signals in the feedback system are from each other, the lower is the directed information between them (measuring distance from starting to finishing sequence and in the direction of cause and effect). 
Thus, post processing signals within a feedback loop, cannot increase the information, which is similar to the open loop case. 
In order to obtain this results, we considered arbitrary causal systems that are interconnected in a feedback loop, with arbitrarily distributed signals.
We were able to overcome the  generally non-trivial  dependencies between the signals in such scenario by establishing a family of useful Markov chains that conditionally decouple the sequences in the system. 
These Markov chains are useful by themselves for studies involving interconnected systems.
We further used the Markov chains to derive a number of fundamental information inequalities that are applicable to signals that are entirely within feedback loops or where some signals are inside and others outside the loop. 
With the use of these inequalities, we were able to show that the conventional notion of channel capacity is not adequate for \emph{in-the-loop} communications. 
Instead, we provided a new notion of in-the-loop channel capacity, and demonstrated a special case, where the new notion of in-the-loop feedback capacity was achievable. 
As an additional application of our results, wediscussed how they allow one to generalize two known fundamental inequalities in networked control involving directed information. 
We are confident that our  analysis provides useful insights to understand and think about information flows in single-loop feedback systems, and that our results will  serve as a toolbox for research in, e.g., networked control systems or communications within  a feedback loop.

%%%%%%%%%%%%%%%%%%%%%%%%%%%%%%%%%%%%%%%%%%
%% Optional
\appendixtitles{yes} % Leave argument "no" if all appendix headings stay EMPTY (then no dot is printed after "Appendix A"). If the appendix sections contain a heading then change the argument to "yes".
\appendixstart
\appendix
\section{A Fundamental Lemma}\label{sec:fl}

\begin{lem}\label{lem:not_so_obvious}
 In the system shown in Fig.~\ref{fig:2systems}, the exogenous signals $\rvar,\rvaq$ are mutually independent and $\Ssp_{1},\Ssp_{2}$ are deterministic (possibly time-varying) causal measurable functions characterized by
$\rvay^{i}=\Ssp_{1}(\rvar^{i},\rvau^{i})$, 
$\rvau^{i}=\Ssp_{2}(\rvaq^{i},\rvay^{i-1})$, $\forall i\in\set{1,\ldots}$, with $\rvay_{0}=y_{0}$ (deterministic).
For this system,  and 
for every $0\leq j \leq i\leq k$ such that $i-j\leq 1 $ and $i\geq 1$,
the following Markov chain holds
\begin{align}
 \rvar^{k}\longleftrightarrow
(\rvau^{i},\rvay^{j})
\longleftrightarrow
\rvaq^{k},\fspace \forall k\in\Nl.
\end{align}
\end{lem}
\begin{figure}[htpb]
\centering
 \input{2systems.pstex_t}
\caption{Two arbitrary causal systems $\Ssp_{1}, \Ssp_{2}$ interconnected in a feedback loop. 
The exogenous signals $\rvar,\rvaq$ are mutually independent.}
\label{fig:2systems}
\end{figure}
\begin{proof}
Let $\Rev,\Qev,\Uev,\Yev$ be the event spaces of $\rvar^{k},\rvaq^{k},\rvau^{i},\rvay^{j}$, respectively.
Since 
$\rvay^{j}=\Ssp_{1}(\rvar^{j},\rvau^{j})$ 
and
$\rvau^{i}=\Ssp_{2}(\rvaq^{i},\rvay^{i-1})$ 
are deterministic measurable functions, it follows that for every possible pair of events 
$\Usp\in\Uev$, 
$\Ysp\in\Yev$,
the preimage sets 
$\Rsp_{\Usp,\Ysp}\eq\set{r^{k}: \Ssp_{1}(r^{j},u^{j})\in\Ysp , u^{i}\in\Usp}$ 
and
$\Qsp_{\Usp,\Ysp}\eq\set{q^{k}: \Ssp_{2}(q^{i},y^{i-1})\in\Usp, y^{j}\in\Ysp }$ 
are also deterministic and belong to $\Rev$ and $\Qev$, respectively.
Thus, 
$(\rvau^{i},\rvay^{j})\in\Usp\times\Ysp\iff
(\rvar^{k}\in\Rsp_{\Usp,\Ysp}, \rvaq^{k}\in\Qsp_{\Usp,\Ysp})$.
This means that for every pair of events $R\in\Rev,Q\in\Qev$,
\begin{align*}
\Pr\set{\rvar^{k}\in R &, \rvaq^{k}\in Q|\rvay^{j}\in\Ysp,\rvau^{i}\in\Usp} 
\\&
\overset{(a)}{=} 
\Pr\set{\rvar^{k}\in R , \rvaq^{k}\in Q|\rvar^{k}\in\Rsp_{\Usp,\Ysp}\,,\, \rvaq^{k}\in\Qsp_{\Usp,\Ysp} }
\\&
\overset{(b)}{=} 
\frac{
\Pr\set{
\rvar^{k}\in R\cap \Rsp_{\Usp,\Ysp}\ ,\ 
\rvaq^{k}\in Q\cap \Qsp_{\Usp,\Ysp} }
}
{\Pr\set{\rvar^{k}\in\Rsp_{\Usp,\Ysp}\,,\, \rvaq^{k}\in\Qsp_{\Usp,\Ysp} }}
\\&
\overset{(c)}{=} 
\frac{\Pr\set{\rvar^{k}\in R\cap \Rsp_{\Usp,\Ysp}}}
{\Pr\set{\rvar^{k}\in\Rsp_{\Usp,\Ysp}}}
\cdot
\frac{\Pr\set{\rvaq^{k}\in Q\cap \Qsp_{\Usp,\Ysp} }}
{\Pr\set{\rvaq^{k}\in\Qsp_{\Usp,\Ysp} }}
\\&
\overset{\hphantom{(d)}}{=} 
\frac{\Pr\set{\rvar^{k}\in R\cap \Rsp_{\Usp,\Ysp}}\Pr\set{\rvaq^{k}\in\Qsp_{\Usp,\Ysp}} }
{\Pr\set{\rvar^{k}\in\Rsp_{\Usp,\Ysp}}\Pr\set{\rvaq^{k}\in\Qsp_{\Usp,\Ysp}} }
\cdot
\frac{\Pr\set{\rvaq^{k}\in Q\cap \Qsp_{\Usp,\Ysp} }\Pr\set{\rvar^{k}\in\Rsp_{\Usp,\Ysp}}}
{\Pr\set{\rvaq^{k}\in\Qsp_{\Usp,\Ysp} }\Pr\set{\rvar^{k}\in\Rsp_{\Usp,\Ysp}} }
\\&
\overset{(d)}{=} 
\frac{\Pr\set{\rvar^{k}\in R\cap \Rsp_{\Usp,\Ysp}\ ,\ \rvaq^{k}\in\Qsp_{\Usp,\Ysp}} }
{\Pr\set{\rvar^{k}\in\Rsp_{\Usp,\Ysp}\ ,\ \rvaq^{k}\in\Qsp_{\Usp,\Ysp}} }
\cdot
\frac{\Pr\set{\rvaq^{k}\in Q\cap \Qsp_{\Usp,\Ysp} \ ,\ \rvar^{k}\in\Rsp_{\Usp,\Ysp}}}
{\Pr\set{\rvaq^{k}\in\Qsp_{\Usp,\Ysp}\ ,\ \rvar^{k}\in\Rsp_{\Usp,\Ysp}} }
\\&
\overset{(e)}{=} 
\Pr\set{\rvar^{k}\in R\ |\ \rvar^{k}\in\Rsp_{\Usp,\Ysp}\ ,\ \rvaq^{k}\in\Qsp_{\Usp,\Ysp}} 
\cdot
\Pr\set{\rvaq^{k}\in Q \ |\   \rvaq^{k}\Qsp_{\Usp,\Ysp} \ ,\ \rvar^{k}\in\Rsp_{\Usp,\Ysp}}
\\&
\overset{(f)}{=} 
\Pr\set{\rvar^{k}\in R\ |\ \rvay^{j}\in\Ysp\ ,\ \rvau^{i}\in\Usp} 
\cdot
\Pr\set{\rvaq^{k}\in Q \ |\   \rvay^{j}\in\Ysp, \rvau^{i} \in\Usp}
%---------------------------------
% %
% \overset{\hphantom{(a)}}{=} 
% \Pr\set{\rvar^{k}\in R |\rvar^{k}\in\Rsp_{\Usp,\Ysp}\,,\, \rvaq^{k}\in\Qsp_{\Usp,\Ysp} }
% \Pr\set{\rvaq^{k}\in Q |\rvar^{k}\in (\Rsp_{\Usp,\Ysp}\cap R)\,,\, \rvaq^{k}\in\Qsp_{\Usp,\Ysp} }
% \\&
% \overset{(a)}{=}
% \Pr\set{\rvar^{k}\in R |\rvar^{k}\in\Rsp_{\Usp,\Ysp}}
% \Pr\set{\rvaq^{k}\in Q |\rvaq^{k}\in\Qsp_{\Usp,\Ysp} }
% \\&
% \overset{\hphantom{(a)}}{=} 
% \Pr\set{\rvar^{k}\in R |\rvay^{j}\in\Ysp,\rvau^{i}\in\Usp}
% \Pr\set{\rvaq^{k}\in Q |\rvay^{j}\in\Ysp,\rvau^{i}\in\Usp},
\end{align*}
where $(a)$ and $(f)$ follow because of the equivalence between the events 
$(\rvay^{j}\in\Ysp,\rvau^{i}\in\Usp)$ and 
$(\rvar^{k}\in\Rsp_{\Usp,\Ysp},\rvaq^{k}\in\Qsp_{\Usp,\Ysp})$,
$(b)$ and $(e)$ follow from Bayes rule,
and $(c)$ and $(d)$
are true because $\rvar^{k}\Perp \rvaq^{k}$.
This completes the proof.
\end{proof}

\section{Proofs}\label{sec:proofs}

\begin{proof}[Proof of Theorem~\ref{thm:full_D-DPI}]\label{prf:full_D-DPI}
If  $(\rvaq,\rvas)\Perp (\rvar,\rvap)$ and $\rvaq\Perp\rvas$, then~\eqref{eq:full_Dpi} follows by applying Theorem~\ref{thm:DPI_dir_dir} and then Theorem~\ref{thm:finally}.
 If $(\rvap,\rvas)\Perp (\rvar,\rvaq)$ and $\rvap\Perp\rvas$, then one arrives to~\eqref{eq:full_Dpi} by applying Theorem~\ref{thm:finally} followed by Theorem~\ref{thm:DPI_dir_dir}.

To prove the second part, notice that
\begin{align}
  I(\rvax^{k}\to\rvay^{k}\parallel \rvaq^{k})
=
I(\rvax^{k}\to\rvay^{k}| \rvaq^{k})\label{eq:dag}
\end{align}
which follows since $\rvax^{i-d_{3}(i)},\rvay^{i}$ are deterministic functions of $(\rvar^{k},\rvap^{k},\rvas^{i},\rvaq^{i})$ and
$\rvaq_{i+1}^{k}\leftrightarrow \rvaq^{i}\leftrightarrow (\rvar^{k},\rvap^{k},\rvas^{i})$,
a Markov chain that results from combining $\rvaq_{i+1}^{k}\leftrightarrow \rvaq^{i}\leftrightarrow \rvas^{i}$ with
$(\rvaq^{k},\rvas^{k})\Perp(\rvar^{k},\rvap^{k})$.

On the other hand, the fact that $(\rvar^{k},\rvap^{k})\Perp(\rvaq^{k},\rvas^{k})$ allows one to obtain from Theorem~\ref{thm:main} that
\begin{align}\label{eq:dem}
 I(\rvax^{k}\to\rvay^{k}|\rvaq^{k}) 
=
I(\rvar^{k},\rvap^{k};\rvay^{k}|\rvaq^{k}).
\end{align}
But
\begin{align}
I(\rvar^{k},\rvap^{k};\rvay^{k}|\rvaq^{k})
&
\overset{\eqref{eq:chainrule_I}}{=}
I(\rvar^{k},\rvap^{k}\,;\, \rvau^{k},\rvay^{k}| \rvaq^{k})
-
I(\rvar^{k},\rvap^{k};\rvau^{k}|\rvaq^{k},\rvay^{k})
\nonumber
\\&
\overset{(a)}{=}
I(\rvar^{k},\rvap^{k}\,;\, \rvau^{k},\rvay^{k}| \rvaq^{k})
\nonumber
\\&
\overset{\eqref{eq:chainrule_I}}{=}
I(\rvar^{k},\rvap^{k}\,;\, \rvau^{k},\rvay^{k}, \rvaq^{k})
-
I(\rvar^{k},\rvap^{k}\,;\,\rvaq^{k})
\nonumber
\\&
\overset{(b)}{=}
I(\rvar^{k},\rvap^{k}\,;\, \rvau^{k},\rvay^{k}, \rvaq^{k})
\nonumber
\\&
\overset{\eqref{eq:chainrule_I}}{=}
I(\rvar^{k},\rvap^{k}\,;\, \rvau^{k})
+
I(\rvar^{k},\rvap^{k}; \rvay^{k},\rvaq^{k} | \rvau^{k})
\label{eq:yfrp}
 \end{align}
where
$(a)$ is due to the fact that 
$\rvau^{k}$ is a deterministic function of $\rvaq^{k},\rvay^{k}$.
Equality $(b)$ holds if and only if $(r,\rvap)\Perp \rvaq$.
The fact that $(\rvar^{k},\rvap^{k})\Perp(\rvaq^{k},\rvas^{k})$ allows one to obtain from Theorem~\ref{thm:main} that
$I(\rvax^{k}\to\rvau^{k})=I(\rvar^{k},\rvap^{k};\rvau^{k})$.
Substituting this in~\eqref{eq:yfrp} and then  into~\eqref{eq:dem} and the latter into~\eqref{eq:dag},  we obtain
$
  I(\rvax^{k}\to\rvay^{k}\parallel \rvaq^{k})
\geq 
I(\rvax^{k}\to\rvau^{k})
$, which combined  with Theorem~\ref{thm:DPI_dir_dir} yields~\eqref{eq:silderok}.
This completes the proof.
\end{proof}

\begin{proof}[Proof of Theorem~\ref{thm:from_splitting_more_precise}]
Apply the chain-rule identity~\eqref{eq:chainrule_I} to the \emph{right-hand side} (RHS) of~\eqref{eq:main_thm} to obtain
\begin{align}\label{eq:Itheta_to_sum}
  I(\theta^{k};\rvay^{k})
=
  I(\rvap^{k},\rvaq^{k},\rvar^{k};\rvay^{k})
=
  I(\rvap^{k},\rvaq^{k};\rvay^{k}|\rvar^{k})
+
I(\rvar^{k};\rvay^{k}).
\end{align}
Now, applying~\eqref{eq:chainrule_I} twice, one can express the term  
$I(\rvap^{k},\rvaq^{k};\rvay^{k}|\rvar^{k})$
as follows:
\begin{equation}\label{eq:estaotra}
 \begin{split}
 I(\rvap^{k},\rvaq^{k};\rvay^{k}|\rvar^{k})
&=
I(\rvap^{k},\rvaq^{k}\,;\,\rvay^{k},\rvar^{k})
-
I(\rvap^{k},\rvaq^{k};\rvar^{k})
=
I(\rvap^{k},\rvaq^{k}\,;\,\rvay^{k},\rvar^{k})
\\&
=
I(\rvap^{k},\rvaq^{k};\rvay^{k})
+
I(\rvap^{k},\rvaq^{k};\rvar^{k}|\rvay^{k}),
% \\&
% \geq 
% I(\rvap^{k},\rvaq^{k};\rvay^{k})
\end{split}
\end{equation}
where the second equality follows since $(\rvap^{k},\rvaq^{k})\Perp \rvar^{k}$. 
The result then follows directly by combining~\eqref{eq:estaotra} with~\eqref{eq:Itheta_to_sum} and~\eqref{eq:main_thm}.
\end{proof}

%-----------------------------------------------------
\begin{proof}[Proof of Theorem~\ref{thm:three_full_loops}]
Since $\rvaq\Perp(\rvar,\rvap,\rvas)$, 
 \begin{align}
 I(\rvax^{k}\to\rvay^{k})
&
\overset{(a)}{=}
I(\rvax^{k}\to\rvau^{k}) + I(\rvaq^{k};\rvay^{k}) + I(\rvar^{k},\rvap^{k};\rvay^{k}|\rvau^{k})
\\
&
\overset{(b)}{=}
I(\rvar^{k},\rvap^{k};\rvau^{k}) + I(\rvaq^{k};\rvay^{k}) + I(\rvar^{k},\rvap^{k};\rvay^{k}|\rvau^{k})
\\
&
\overset{(c)}{=}
I(\rvar^{k};\rvau^{k}) + I(\rvap^{k};\rvau^{k}|\rvar^{k}) + I(\rvaq^{k};\rvay^{k}) + I(\rvar^{k},\rvap^{k};\rvay^{k}|\rvau^{k}),
\label{eq:laotraultima}
\end{align}
where $(a)$ is due to Theorem~\ref{thm:finally}, $(b)$ follows from Theorem~\ref{thm:main} and the fact that $(\rvas,\rvaq)\Perp (\rvar,\rvap)$ and $(c)$ from the chain rule of mutual information.
For the second term on the RHS of the last equation, we have 
\begin{align}
 I(\rvap^{k};\rvau^{k}|\rvar^{k})
&
\overset{(a)}{=}
I(\rvap^{k};\rvau^{k}|\rvar^{k}) + I(\rvap^{k};\rvar^{k})
=
I(\rvap^{k};\rvar^{k},\rvau^{k})
\\&\overset{(b)}{=}
I(\rvap^{k};\rvar^{k},\rvau^{k},\rvae^{k})
-
I(\rvap^{k};\rvae^{k}|\rvar^{k},\rvau^{k})
\\&\overset{(c)}{=}
I(\rvap^{k};\rvar^{k},\rvau^{k},\rvae^{k})
\\&\overset{(d)}{=}
I(\rvap^{k};\rvae^{k})
+
I(\rvap^{k};\rvar^{k},\rvau^{k}|\rvae^{k})
\\&\overset{(e)}{=}
I(\rvap^{k};\rvae^{k})
+
I(\rvap^{k};\rvau^{k}|\rvae^{k})
+
I(\rvap^{k};\rvar^{k}|\rvau^{k},\rvae^{k})
\\&\overset{(f)}{=}
I(\rvap^{k};\rvae^{k})
+
I(\rvap^{k};\rvau^{k}|\rvae^{k}),
\label{eq:ultima_linea}
\end{align}
where $(a)$ holds since $\rvar\Perp\rvap$, $(b)$, $(d)$ and $(e)$ stem from the chain rule of mutual information~\eqref{eq:chainrule_I}, and $(c)$ is a consequence of the fact that $\rvae^{k}=\Ssp_{1}(\rvau^{k-d_{1}(k)},\rvar^{k})$.
Finally, $(f)$ is due to the Markov chain 
$
\rvar^{k}
\leftrightarrow
(\rvau^{k},\rvae^{k})
\leftrightarrow
\rvap^{k}
$,
which holds because $\rvar\Perp(\rvap,\rvas,\rvaq)$ as a consequence of Lemma~\ref{lem:not_so_obvious} in the appendix (see also Fig.~\ref{fig:diagramas}-(a)).
Substitution of~\eqref{eq:ultima_linea} into~\eqref{eq:laotraultima} yields~\eqref{eq:nice}, thereby completing the proof.
\end{proof}

\begin{proof}[Proof of Theorem~\ref{thm:DPI_dir_dir}]
Since $(\rvap,\rvaq,\rvar)\Perp \rvas$, we can apply~\eqref{eq:la_de_Massey} (where now $(\rvaq,\rvar)$ plays 
the role of $\rvar$), 
and obtain
\begin{align}
 I(\rvax^{k}\to \rvay^{k})\geq I(\rvaq^{k},\rvar^{k};\rvay^{k}).
\end{align}
Now, we apply Theorem~\ref{thm:main}, which gives  
\begin{align}
 I(\rvaq^{k},\rvar^{k};\rvay^{k}) \geq  I(\rvae^{k}\to \rvay^{k}),
\end{align}
completing the proof.
\end{proof}

\begin{proof}[Proof of Theorem~\ref{thm:finally}]
% Applying Theorem~\ref{thm:main}, since $(\rvar,\rvap) \Perp(\rvas,\rvaq)$, 
% %
% \begin{align}\label{eq:qtoy=r,u}
%  I(\rvax^{k}\to \rvau^{k}) = I(\rvar^{k},\rvap^{k}\,;\, \rvau^{k}).
% \end{align}
% %
% For the other directed information,
We have that  
\begin{align}
 I(\rvax^{k}\to \rvay^{k})
&
\overset{(a)}{=}
I(\rvar^{k},\rvap^{k},\rvaq^{k}\,;\, \rvay^{k})
\nonumber
\\&
\overset{\eqref{eq:chainrule_I}}{=}
I(\rvaq^{k}\,;\, \rvay^{k})
+
I(\rvar^{k},\rvap^{k}\,;\, \rvay^{k}| \rvaq^{k})
\label{eq:lamisma_I}
\\&
\overset{\eqref{eq:yfrp}}{=}
I(\rvar^{k},\rvap^{k}\,;\, \rvau^{k})
+
I(\rvar^{k},\rvap^{k}; \rvay^{k},\rvaq^{k} | \rvau^{k})
\nonumber
\\&
\overset{\eqref{eq:chainrule_I}}{=}
I(\rvaq^{k}\,;\, \rvay^{k})
+
I(\rvar^{k},\rvap^{k}\,;\, \rvau^{k})
+
I(\rvar^{k},\rvap^{k}; \rvay^{k}| \rvau^{k})
 +
 I(\rvar^{k},\rvap^{k};\rvaq^{k}|\rvau^{k},\rvay^{k})
\label{eq:penultima}
\\&
\overset{(b)}{\geq}
I(\rvaq^{k}\,;\, \rvay^{k})
+
I(\rvax^{k}\to \rvau^{k})
+
I(\rvar^{k},\rvap^{k}; \rvay^{k}| \rvau^{k})
 +
 I(\rvar^{k},\rvap^{k};\rvaq^{k}|\rvau^{k},\rvay^{k})
\\&
\overset{(c)}{\geq}
I(\rvaq^{k}\,;\, \rvay^{k})
+
I(\rvax^{k}\to \rvau^{k})
+
I(\rvar^{k},\rvap^{k}; \rvay^{k}| \rvau^{k}),\label{eq:ladeabajo}
\end{align}
where 
$(a)$ follows from Theorem~\ref{thm:main} and the  assumption $(\rvar,\rvap,\rvaq) \Perp \rvas$, 
$(b)$ is from Theorem~\ref{thm:main}, with equality iff $(\rvaq,\rvas)\Perp(\rvar,\rvap)$,
and
from Lemma~\ref{lem:not_so_obvious} (in the appendix),
$(c)$ turns into equality if $\rvaq\Perp (\rvar,\rvap,\rvas)$. 
This completes the proof.
\end{proof}

%%%%%%%%%%%%%%%%%%%%%%%%%%%%%%%%%%%%%%%%%%
% \authorcontributions{For research articles with several authors, a short paragraph specifying their individual contributions must be provided. The following statements should be used ``Conceptualization, X.X. and Y.Y.; methodology, X.X.; software, X.X.; validation, X.X., Y.Y. and Z.Z.; formal analysis, X.X.; investigation, X.X.; resources, X.X.; data curation, X.X.; writing---original draft preparation, X.X.; writing---review and editing, X.X.; visualization, X.X.; supervision, X.X.; project administration, X.X.; funding acquisition, Y.Y. All authors have read and agreed to the published version of the manuscript.'', please turn to the  \href{http://img.mdpi.org/data/contributor-role-instruction.pdf}{CRediT taxonomy} for the term explanation. Authorship must be limited to those who have contributed substantially to the work~reported.}
% 
% \funding{
% % This research received no external funding
% This research was funded by NAME OF FUNDER grant number XXX}
% 
% 
% 
% \acknowledgments{In this section you can acknowledge any support given which is not covered by the author contribution or funding sections. This may include administrative and technical support, or donations in kind (e.g., materials used for experiments).}
% 
% \conflictsofinterest{The authors declare no conflict of interest.} 

%%%%%%%%%%%%%%%%%%%%%%%%%%%%%%%%%%%%%%%%%%
%% Only for journal Encyclopedia
%\entrylink{The Link to this entry published on the encyclopedia platform.}

%%%%%%%%%%%%%%%%%%%%%%%%%%%%%%%%%%%%%%%%%%

\reftitle{References}

% Please provide either the correct journal abbreviation (e.g. according to the "List of Title Word Abbreviations" http://www.issn.org/services/online-services/access-to-the-ltwa/) or the full name of the journal.

% \externalbibliography{yes}
% \bibliography{\BibPath/IEEEabrv,%

\begin{thebibliography}{999}

\bibitem[Cover and Thomas(2006)]{covtho06}
Cover, T.M.; Thomas, J.A.
\newblock {\em Elements of Information Theory}, 2nd ed.; Wiley-Interscience:
  Hoboken, N.J,  2006.

\bibitem[{Salek} \em{et~al.}(2019){Salek}, {Cadamuro}, {Kammerlander}, and
  {Wiesner}]{salcad19}
{Salek}, S.; {Cadamuro}, D.; {Kammerlander}, P.; {Wiesner}, K.
\newblock Quantum Rate-Distortion Coding of Relevant Information.
\newblock {\em {IEEE} Transactions on Information Theory} {\bf 2019}, {\em
  65},~2603--2613.
\newblock
  doi:{\changeurlcolor{black}\href{https://doi.org/10.1109/TIT.2018.2878412}{\detokenize{10.1109/TIT.2018.2878412}}}.

\bibitem[{Lindenstrauss} and {Tsukamoto}(2018)]{lintsu18}
{Lindenstrauss}, E.; {Tsukamoto}, M.
\newblock From Rate Distortion Theory to Metric Mean Dimension: Variational
  Principle.
\newblock {\em {IEEE} Transactions on Information Theory} {\bf 2018}, {\em
  64},~3590--3609.
\newblock
  doi:{\changeurlcolor{black}\href{https://doi.org/10.1109/TIT.2018.2806219}{\detokenize{10.1109/TIT.2018.2806219}}}.

\bibitem[{Yang} \em{et~al.}(2017){Yang}, {Grover}, and {Kar}]{yangro17}
{Yang}, Y.; {Grover}, P.; {Kar}, S.
\newblock Rate Distortion for Lossy In-Network Linear Function Computation and
  Consensus: Distortion Accumulation and Sequential Reverse Water-Filling.
\newblock {\em {IEEE} Transactions on Information Theory} {\bf 2017}, {\em
  63},~5179--5206.
\newblock
  doi:{\changeurlcolor{black}\href{https://doi.org/10.1109/TIT.2017.2710059}{\detokenize{10.1109/TIT.2017.2710059}}}.

\bibitem[Derpich and {\O}stergaard(2012)]{derost12}
Derpich, M.S.; {\O}stergaard, J.
\newblock Improved upper bounds to the causal quadratic rate-distortion
  function for {G}aussian stationary sources.
\newblock {\em {IEEE} Transactions on Information Theory} {\bf 2012}, {\em
  58},~3131--3152.

\bibitem[{Ramakrishnan} \em{et~al.}(2021){Ramakrishnan}, {Iten}, {Scholz}, and
  {Berta}]{ramite21}
{Ramakrishnan}, N.; {Iten}, R.; {Scholz}, V.B.; {Berta}, M.
\newblock Computing Quantum Channel Capacities.
\newblock {\em {IEEE} Transactions on Information Theory} {\bf 2021}, {\em
  67},~946--960.
\newblock
  doi:{\changeurlcolor{black}\href{https://doi.org/10.1109/TIT.2020.3034471}{\detokenize{10.1109/TIT.2020.3034471}}}.

\bibitem[Song \em{et~al.}(2020)Song, Zhang, Kadhe, Bakshi, and Jaggi]{sonzha20}
Song, J.; Zhang, Q.; Kadhe, S.; Bakshi, M.; Jaggi, S.
\newblock Stealthy Communication Over Adversarially Jammed Multipath Networks.
\newblock {\em {IEEE} Transactions on Information Theory} {\bf 2020}, {\em
  68},~7473--7484.
\newblock
  doi:{\changeurlcolor{black}\href{https://doi.org/10.1109/TCOMM.2020.3022785}{\detokenize{10.1109/TCOMM.2020.3022785}}}.

\bibitem[{Makur}(2020)]{makur-20}
{Makur}, A.
\newblock Coding Theorems for Noisy Permutation Channels.
\newblock {\em {IEEE} Transactions on Information Theory} {\bf 2020}, {\em
  66},~6723--6748.
\newblock
  doi:{\changeurlcolor{black}\href{https://doi.org/10.1109/TIT.2020.3009468}{\detokenize{10.1109/TIT.2020.3009468}}}.

\bibitem[Kostina and Verd\'u(2013)]{kosver13}
Kostina, V.; Verd\'u, S.
\newblock Lossy joint source-channel coding in the finite blocklength regime.
\newblock {\em {IEEE} Transactions on Information Theory} {\bf 2013}, {\em
  59},~2545--2575.

\bibitem[{Huang} and {Narayanan}(2012)]{huanar12}
{Huang}, Y.; {Narayanan}, K.R.
\newblock Joint Source-Channel Coding with Correlated Interference.
\newblock {\em {IEEE} Trans. Commun.} {\bf 2012}, {\em 60},~1315--1327.
\newblock
  doi:{\changeurlcolor{black}\href{https://doi.org/10.1109/TCOMM.2012.032012.110185}{\detokenize{10.1109/TCOMM.2012.032012.110185}}}.

\bibitem[{Steinberg} and {Merhav}(2006)]{stemer06}
{Steinberg}, Y.; {Merhav}, N.
\newblock On hierarchical joint source-channel coding with degraded side
  information.
\newblock {\em {IEEE} Transactions on Information Theory} {\bf 2006}, {\em
  52},~886--903.
\newblock
  doi:{\changeurlcolor{black}\href{https://doi.org/10.1109/TIT.2005.864423}{\detokenize{10.1109/TIT.2005.864423}}}.

\bibitem[Massey(1990)]{massey90}
Massey, J.L.
\newblock Causality, feedback and directed information.
\newblock  Proc.~ Intl.~Symp.~Inf.~Theory and its Appl.; ,  1990; pp. 303--305.

\bibitem[Kramer(1998)]{kramer98}
Kramer, G.
\newblock Directed information for channels with feedback.
\newblock PhD thesis, Swiss federal institute of technology,  1998.

\bibitem[Tatikonda and Mitter(2009)]{tatmit09}
Tatikonda, S.; Mitter, S.
\newblock The Capacity of Channels With Feedback.
\newblock {\em {IEEE} Transactions on Information Theory} {\bf 2009}, {\em
  55},~323--349.
\newblock
  doi:{\changeurlcolor{black}\href{https://doi.org/10.1109/TIT.2008.2008147}{\detokenize{10.1109/TIT.2008.2008147}}}.

\bibitem[{Li} and {Elia}(2011)]{li-eli11}
{Li}, C.; {Elia}, N.
\newblock The Information Flow and Capacity of Channels with Noisy Feedback.
\newblock {\em Submitted to {IEEE} Transactions on Information Theory} {\bf
  2011},  \href{http://xxx.lanl.gov/abs/1108.2815}{{\normalfont
  [arXiv:cs.IT/1108.2815]}}.

\bibitem[Tatikonda(2000)]{tatiko00}
Tatikonda, S.C.
\newblock Control under Communication Constraints.
\newblock PhD thesis, Department of Electrical Engineering and Computer
  Science, Massachusetts Institute of Technology, Cambridge, MA,  2000.

\bibitem[Martins and Dahleh(2005)]{mardah05}
Martins, N.C.; Dahleh, M., M.A.
\newblock Fundamental limitations of performance in the presence of finite
  capacity feedback.
\newblock  Proc.~American Control Conf.,  2005.

\bibitem[Martins and Dahleh(2008)]{mardah08}
Martins, N.; Dahleh, M.
\newblock Feedback control in the presence of noisy Channels: ``{B}ode-like''
  fundamental limitations of performance.
\newblock {\em {IEEE} Transactions on Automatic Control} {\bf 2008}, {\em
  53},~1604--1615.
\newblock
  doi:{\changeurlcolor{black}\href{https://doi.org/10.1109/TAC.2008.929361}{\detokenize{10.1109/TAC.2008.929361}}}.

\bibitem[Silva \em{et~al.}(2011)Silva, Derpich, and {\O}stergaard]{silder11}
Silva, E.I.; Derpich, M.S.; {\O}stergaard, J.
\newblock A framework for control system design subject to average data-rate
  constraints.
\newblock {\em {IEEE} Transactions on Automatic Control} {\bf 2011}, {\em
  56},~1886--1899.

\bibitem[Silva \em{et~al.}(2010)Silva, Derpich, and {\O}stergaard]{silder10}
Silva, E.I.; Derpich, M.S.; {\O}stergaard, J.
\newblock On the Minimal Average Data-Rate That Guarantees a Given Closed Loop
  Performance Level.
\newblock  Proc.~2nd IFAC Workshop on Distributed Estimation and Control in
  Networked Systems, NECSYS; ,  2010; pp. 67--72.

\bibitem[Silva \em{et~al.}(2011)Silva, Derpich, and {\O}stergaard]{silder11b}
Silva, E.I.; Derpich, M.S.; {\O}stergaard, J.
\newblock An achievable data-rate region subject to a stationary performance
  constraint for {LTI} plants.
\newblock {\em {IEEE} Transactions on Automatic Control} {\bf 2011}, {\em
  56},~1968--1973.

\bibitem[Tanaka \em{et~al.}(2018)Tanaka, Esfahani, and Mitter]{tanesf18}
Tanaka, T.; Esfahani, P.M.; Mitter, S.K.
\newblock {LQG} Control With Minimum Directed Information: Semidefinite
  Programming Approach.
\newblock {\em {IEEE} Transactions on Automatic Control} {\bf 2018}, {\em
  63},~37--52.
\newblock
  doi:{\changeurlcolor{black}\href{https://doi.org/10.1109/TAC.2017.2709618}{\detokenize{10.1109/TAC.2017.2709618}}}.

\bibitem[Quinn \em{et~al.}(2011)Quinn, Coleman, Kiyavash, and
  Hatsopoulos]{quicol11}
Quinn, C.; Coleman, T.; Kiyavash, N.; Hatsopoulos, N.
\newblock Estimating the directed information to infer causal relationships in
  ensemble neural spike train recordings.
\newblock {\em Journal of Computational Neuroscience} {\bf 2011}, {\em
  30},~17--44.
\newblock
  doi:{\changeurlcolor{black}\href{https://doi.org/10.1007/s10827-010-0247-2}{\detokenize{10.1007/s10827-010-0247-2}}}.

\bibitem[Permuter \em{et~al.}(2011)Permuter, Kim, and Weissman]{perkim11}
Permuter, H.H.; Kim, Y.H.; Weissman, T.
\newblock Interpretations of directed information in portfolio theory, data
  Compression, and hypothesis testing.
\newblock {\em {IEEE} Transactions on Information Theory} {\bf 2011}, {\em
  57},~3248--3259.

\bibitem[Derpich and {\O}stergaard(2021)]{derost21b}
Derpich, M.S.; {\O}stergaard, J.
\newblock Comments on "A Framework for Control System Design Subject to Average
  Data-Rate Constraints".
\newblock {\em Submitted to IEEE Transactions on Automatic Control, avail. from
  arxiv.org} {\bf 2021}.

\bibitem[Massey and Massey(2005)]{masmas05}
Massey, J.; Massey, P.
\newblock Conservation of mutual and directed information.
\newblock  Proc.~IEEE Int.~Symp.~Information Theory,  2005, pp. 157--158.
\newblock
  doi:{\changeurlcolor{black}\href{https://doi.org/10.1109/ISIT.2005.1523313}{\detokenize{10.1109/ISIT.2005.1523313}}}.

\bibitem[Kim and Kim(2008)]{kim-yh08}
Kim, Y.H.; Kim, Y.H.
\newblock A Coding Theorem for a Class of Stationary Channels With Feedback.
\newblock {\em {IEEE} Transactions on Information Theory} {\bf 2008}, {\em
  54},~1488--1499.
\newblock
  doi:{\changeurlcolor{black}\href{https://doi.org/10.1109/TIT.2008.917685}{\detokenize{10.1109/TIT.2008.917685}}}.

\bibitem[Zamir \em{et~al.}(2008)Zamir, Kochman, and Erez]{zamkoc08}
Zamir, R.; Kochman, Y.; Erez, U.
\newblock Achieving the {G}aussian rate-distortion function by prediction.
\newblock {\em {IEEE} Transactions on Information Theory} {\bf 2008}, {\em
  54},~3354--3364.

\bibitem[Zhang and Sun(2006)]{zhasun06}
Zhang, H.; Sun, Y.X.
\newblock Directed information and mutual information in linear feedback
  tracking systems.
\newblock  Proc. 6-th World Congress on Intelligent Control and Automation,
  2006, pp. 723--727.

\bibitem[Silva \em{et~al.}(2016)Silva, Derpich, {\O}stergaard, and
  Encina]{silder16}
Silva, E.I.; Derpich, M.S.; {\O}stergaard, J.; Encina, M.A.
\newblock A characterization of the minimal average data rate that guarantees a
  given closed-lop performance level.
\newblock {\em {IEEE} Transactions on Automatic Control} {\bf 2016}, {\em
  61},~2171--2186.
\newblock
  doi:{\changeurlcolor{black}\href{https://doi.org/10.1109/TAC.2015.2500658}{\detokenize{10.1109/TAC.2015.2500658}}}.

\bibitem[Derpich \em{et~al.}(2013)Derpich, Silva, and {\O}stergaard]{derost13}
Derpich, M.S.; Silva, E.I.; {\O}stergaard, J.
\newblock Fundamental Inequalities and Identities Involving Mutual and Directed
  Informations in Closed-Loop Systems.
\newblock {\em ArXiv e-prints} {\bf 2013}, {\em abs/1301.6427}.

\bibitem[Shahsavari~Baboukani \em{et~al.}()Shahsavari~Baboukani, Graversen,
  Alickovic, and Østergaard]{shagra20}
Shahsavari~Baboukani, P.; Graversen, C.; Alickovic, E.; {\O}stergaard, J.
\newblock Estimating Conditional Transfer Entropy in Time Series Using Mutual
  Information and Nonlinear Prediction.
\newblock {\em Entropy}, {\em 22}.
\newblock
  doi:{\changeurlcolor{black}\href{https://doi.org/10.3390/e22101124}{\detokenize{10.3390/e22101124}}}.

\bibitem[{Barforooshan} \em{et~al.}(2020){Barforooshan}, {Derpich}, {Stavrou},
  and {Ostergaard}]{barder20}
{Barforooshan}, M.; {Derpich}, M.S.; {Stavrou}, P.A.; {Ostergaard}, J.
\newblock The Effect of Time Delay on the Average Data Rate and Performance in
  Networked Control Systems.
\newblock {\em {IEEE} Transactions on Automatic Control} {\bf 2020}, pp. 1--1.
\newblock
  doi:{\changeurlcolor{black}\href{https://doi.org/10.1109/TAC.2020.3047578}{\detokenize{10.1109/TAC.2020.3047578}}}.

\bibitem[{Baboukani} \em{et~al.}(2021){Baboukani}, {Graversen}, and
  {{\O}stergaard}]{babgra20}
{Baboukani}, P.S.; {Graversen}, C.; {{\O}stergaard}, J.
\newblock Estimation of Directed Dependencies in Time Series Using Conditional
  Mutual Information and Non-linear Prediction.
\newblock  2020 28th European Signal Processing Conference (EUSIPCO),  2021,
  pp. 2388--2392.
\newblock
  doi:{\changeurlcolor{black}\href{https://doi.org/10.23919/Eusipco47968.2020.9287592}{\detokenize{10.23919/Eusipco47968.2020.9287592}}}.

\bibitem[Yeh(2014)]{yeh---14}
Yeh, J.
\newblock {\em Real analysis}, 3rd ed ed.; World Scientific,  2014.

\bibitem[Gray(2011)]{gray--11}
Gray, R.M.
\newblock {\em Entropy and Information Theory}, 2 ed.; Science+Business Media,
  Springer: New York,  2011.

\bibitem[Yeung(2002)]{yeung-02}
Yeung, R.W.
\newblock {\em A first course in {I}nformation {T}heory}; Springer,  2002.

\bibitem[Goodwin \em{et~al.}()Goodwin, Graebe, and Salgado]{gogrsa01}
Goodwin, G.C.; Graebe, S.; Salgado, M.E.
\newblock {\em Control System Design}; Prentice Hall.

\bibitem[Sahai and Mitter(2006)]{sahmit06}
Sahai, A.; Mitter, S.
\newblock The Necessity and Sufficiency of Anytime Capacity for Stabilization
  of a Linear System Over a Noisy Communication Link--Part I: Scalar
  Systems.
\newblock {\em {IEEE} Transactions on Information Theory} {\bf 2006}, {\em
  52},~3369--3395.
\newblock
  doi:{\changeurlcolor{black}\href{https://doi.org/10.1109/TIT.2006.878169}{\detokenize{10.1109/TIT.2006.878169}}}.

\bibitem[{Khina} \em{et~al.}(2019){Khina}, {Gårding}, {Pettersson}, {Kostina},
  and {Hassibi}]{khigar19}
{Khina}, A.; {G\aa{}rding}, E.R.; {Pettersson}, G.M.; {Kostina}, V.; {Hassibi}, B.
\newblock Control Over Gaussian Channels With and Without Source?Channel
  Separation.
\newblock {\em {IEEE} Transactions on Automatic Control} {\bf 2019}, {\em
  64},~3690--3705.
\newblock
  doi:{\changeurlcolor{black}\href{https://doi.org/10.1109/TAC.2019.2912255}{\detokenize{10.1109/TAC.2019.2912255}}}.

\end{thebibliography}
% \BibPath/MDerpich%
% }

%=====================================
% References, variant B: internal bibliography
%=====================================
% \begin{thebibliography}{999}
% \end{thebibliography}

% If authors have biography, please use the format below
%\section*{Short Biography of Authors}
%\bio
%{\raisebox{-0.35cm}{\includegraphics[width=3.5cm,height=5.3cm,clip,keepaspectratio]{Definitions/author1.pdf}}
%{\textbf{Firstname Lastname} Biography of first author}
%
%\bio
%{\raisebox{-0.35cm}{\includegraphics[width=3.5cm,height=5.3cm,clip,keepaspectratio]{Definitions/author2.jpg}}
%{\textbf{Firstname Lastname} Biography of second author}

%%%%%%%%%%%%%%%%%%%%%%%%%%%%%%%%%%%%%%%%%%
\end{document}